\pdfoutput=1
\documentclass[%
    10pt               ,   % Type size     : [10pt,11pt,12pt]
    aps                ,   % Main styling  : [aps|aip]
    prl                ,   % Sub-styling   : [pra|prb|prc|prd|pre|prl|prstab|prstper|rmp] or [apl|bmf|cha|jap|jcp|jmp|rse|pof|pop|rsi]
    reprint            ,   % View style    : [preprint|reprint]
    twocolumn          ,   % Columns       : [onecolumn|twocolumn]
    showpacs           ,   % PACS numbers  : [showpacs|noshowpacs]
    groupedaddress     ,   % Author address: [groupedaddress,superscriptaddress,unsortedaddress]
    footinbib          ,   % Footnotes     : [footinbib|nofootinbib]
    citeautoscript     ,   % Fix superscript citation order
    longbibliography   ,   % Longer biblio : [longbibliography,nolongbibliography]
    notitlepage        ,   % Don't take a whole page for the title
    floatfix,
]{revtex4-2}
\usepackage[T1]{fontenc}
\usepackage[utf8]{inputenc}
\usepackage[english]{babel}
\usepackage{titlesec}
\usepackage{comment}
\usepackage{makecell}
\usepackage{booktabs}
\renewcommand{\arraystretch}{1.2}
\renewcommand{\paragraph}[1]{\textit{#1}---\ignorespaces}
\makeatletter
\def\@bibdataout@rev{%
 \if@filesw
  \immediate\write\@auxout{\string\citation{REVTEX42Control}}%
  \immediate\write\@auxout{\string\citation{apsrev42Control}}%
 \fi
}
\makeatother

\usepackage{mathtools}
\allowdisplaybreaks
\usepackage{amssymb}
\usepackage{braket}
\usepackage{dsfont}
\usepackage{siunitx}
\newcommand{\rma}{{\rm a}}
\newcommand{\rme}{{\rm e}}
\newcommand{\rmd}{{\rm d}}
\newcommand{\rmi}{{\rm i}}

\newcommand{\rmc}{{\rm c}}

\newcommand{\bfr}{{\mathbf{r}}}

\newcommand{\bfu}{{\mathbf{u}}}

\newcommand{\bfn}{{\mathbf{n}}}

\newcommand{\LL}{{\mathcal{L}}}
\newcommand{\DD}{{\mathcal{D}}}
\newcommand{\HH}{{\mathcal{H}}}
\newcommand{\JJ}{{\mathcal{J}}}
\newcommand{\UU}{{\mathcal{U}}}
\newcommand{\vket}{{\rangle\!\rangle}}
\newcommand{\vbra}{{\langle\!\langle}}

\newcommand  {\SYK}   {\mathrm{SYK}}                % SYK
\newcommand  {\SFF}   {\mathrm{SFF}}                % SFF
\newcommand  {\DSFF}  {\mathrm{DSFF}}               % DSFF
\newcommand  {\sSFF}  {\sigma\mathrm{SFF}}          % sigma-SFF
\DeclareMathOperator{\Tr}{Tr}
\DeclareMathOperator{\Imag}{Im}
\DeclareMathOperator{\Real}{Re}

\usepackage[dvipsnames]{xcolor} 
\usepackage{graphicx}
\definecolor{darkred}  {rgb}{0.6,0.05,0.05}
\definecolor{darkgreen}{rgb}{0.05,0.6,0.05}
\definecolor{darkblue} {rgb}{0.05,0.05,0.6}
\usepackage[%hypertex,
    unicode           = true  ,
    plainpages        = false , 
    pdfpagelabels     = true  , 
    bookmarks         = true  ,
    bookmarksnumbered = true  ,
    bookmarksopen     = true  ,
    breaklinks        = true  ,
    backref           = false ,
    colorlinks        = true  ,
    hyperindex        = true  ,
    linktocpage       = true  ,
    hyperfigures      = true
]{hyperref}
\hypersetup{
    linkcolor         = darkred   ,
    urlcolor          = darkblue  ,
    filecolor         = darkgreen ,
    citecolor         = darkgreen ,
    anchorcolor       = darkred   ,
}
\usepackage[poorman,capitalise]{cleveref}
\usepackage{orcidlink}

\begin{document}

\title{Dissipation-induced Sachdev-Ye-Kitaev physics in many-body \texorpdfstring{\\}{} cavity quantum electrodynamics}

\author{Pietro Pacchioni\,\orcidlink{0000-0001-7496-3836}}
\altaffiliation{These authors contributed equally to this work. Correspondence should be addressed to filippo.ferrari@epfl.ch}
\affiliation{Institute of Physics and Center for Quantum Science and Engineering,\\ École Polytechnique Fédérale de Lausanne (EPFL), Lausanne, Switzerland}
\author{Filippo Ferrari\,\orcidlink{0009-0003-6317-0816}}
\altaffiliation{These authors contributed equally to this work. Correspondence should be addressed to filippo.ferrari@epfl.ch}
\affiliation{Institute of Physics and Center for Quantum Science and Engineering,\\ École Polytechnique Fédérale de Lausanne (EPFL), Lausanne, Switzerland}
\author{Vincenzo Savona\,\orcidlink{0000-0002-8984-6584}}
\affiliation{Institute of Physics and Center for Quantum Science and Engineering,\\ École Polytechnique Fédérale de Lausanne (EPFL), Lausanne, Switzerland}
\author{Matteo Seclì\,\orcidlink{0000-0002-9608-096X}}
\affiliation{Institute of Physics and Center for Quantum Science and Engineering,\\ École Polytechnique Fédérale de Lausanne (EPFL), Lausanne, Switzerland}

\date{October 6, 2026}

\begin{abstract}
We show that cavity quantum electrodynamics (QED) devices can realize dissipative Sachdev-Ye-Kitaev (SYK) universality, a paradigmatic setting for quantum chaos in open many-body systems. 
Ultracold fermions with disordered, all-to-all cavity-mediated interactions provide two complementary routes: atomic spontaneous emission in a single-mode cavity and photon leakage from a multimode cavity. 
Strikingly, both converge to the same non-Hermitian random-matrix universality despite originating from integrable and chaotic closed-system limits, respectively. 
In the single-mode cavity, dissipation creates quantum chaos from an integrable Hamiltonian. 
We trace this convergence to a tunable growth in dissipative rank, controlled, respectively, by the Lamb-Dicke parameter and the cavity-mode spacing. 
The integrability-to-chaos crossover is accompanied by a dynamical crossover from long-lived prethermal memory to rapid thermalization, visible in single-atom-resolved densities.
\end{abstract}

\maketitle

\paragraph{Introduction}
Quantum chaos underpins the behavior of a broad class of interacting quantum many-body systems~\cite{dalessio_quantum_2016, haake_quantum_2018}.
Thermalization under unitary time evolution~\cite{deutsch_quantum_1991, srednicki_chaos_1994, abanin_colloquium_2019}, quantum information scrambling~\cite{hayden_black_2007, sekino_fast_2008, shenker_black_2014, maldacena_bound_2016, swingle_measuring_2016}, and universal entanglement spreading~\cite{kim_ballistic_2013, hartman_time_2013, mezei_entanglement_2017, bertini_entanglement_2019} are all rooted in an emergent rigidity in the Hamiltonian spectrum~\cite{bohigas_characterization_1984}, which determines the properties of the Hamiltonian eigenstates~\cite{dalessio_quantum_2016, abanin_colloquium_2019} and, consequently, any aspect of the quantum dynamics. 
At the same time, quantum systems interact with a surrounding environment, resulting in deeply modified properties compared to idealized isolated scenarios~\cite{breuer_theory_2007}.
The interplay between chaotic behavior and dissipative mechanisms defines the field of dissipative quantum chaos (DQC), which is rapidly developing~\cite{akemann_universal_2019, denisov_universal_2019, can_random_2019, hamazaki_universality_2020, sa_spectral_2020, sa_complex_2020,  prasad_dissipative_2022, dahan_classical_2022, kawabata_symmetry_2023, villasenor_breakdown_2024, gupta_quantum_2024, ferrari_dissipative_2025, peyruchat_landauzener_2025, ferrari_chaotic_2025, kruglikov_chaos_2025, zundel_nonintegrability_2025, wold2025experimentaldetectiondissipativequantum, mondal_transient_2026, almeida_dissipation-_2026, sa_2026talktalkdissipativequantum}.

A special class of chaotic quantum systems is the Sachdev-Ye-Kitaev (SYK) type of models~\cite{sachdev_gapless_1993, kitaev_simple_2015, maldacena_remarks_2016, rosenhaus_introduction_2019, chowdhury_sachdev-ye-kitaev_2022}, consisting of $N$ fermions subject to random, all-to-all interactions.
Beyond exhibiting the general hallmarks of quantum chaos outlined above, SYK systems stand out for their connections to conjectured holographic phases of matter and, more broadly, to quantum gravity~\cite{maldacena_remarks_2016, rosenhaus_introduction_2019}.
Dissipative generalizations of the SYK model have become, in turn, canonical strongly correlated toy models of non-Hermitian universality~\cite{garcia-garcia_symmetry_2022, kulkarni_lindbladian_2022, sa_lindbladian_2022, garcia-garcia_keldysh_2023, garcia-garcia_universality_2023}, and their experimental realization would be highly desirable.
In this work, we argue that cavity quantum electrodynamics (QED) with ultracold fermionic atoms~\cite{mivehvar_cavity_2021} and randomized couplings~\cite{sauerwein_engineering_2023, orsi_cavity_2024} gives rise to open SYK physics.
Cavity-mediated SYK interactions were initially proposed for closed-system realizations of the model~\cite{Uhrich2023, Baumgartner2024, Baumgartner2025, solis_single-particle_2026, FerrariEtAl2026}, where complicated time-dependent protocols~\cite{Baumgartner2024} were required, and the inherent dissipation of cavity QED setup was seen as an obstacle, destroying several important features~\cite{Baumgartner2025, FerrariEtAl2026}.
These challenges motivate a different approach: targeting \emph{dissipative} SYK physics from the outset and harnessing, rather than suppressing, the intrinsic dissipation of cavity QED.

\begin{figure}[t]
    \centering
    \includegraphics[width=\columnwidth]{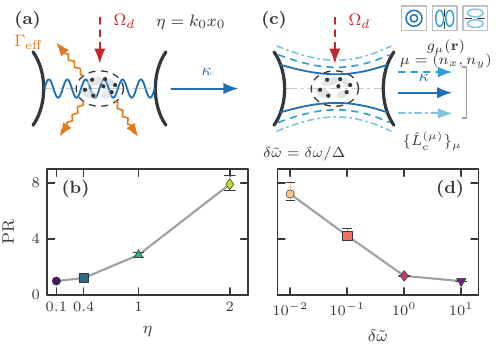}
    \caption{\label{fig:sketch}
    Sketch of the systems and rank structure.
    (a) Single-mode optical cavity dissipating photons at rate $\kappa$ coupled to a driven gas of ultracold fermions subject to spontaneous emission at effective rate $\Gamma_{\rm eff}$.
    The features of spontaneous emission are controlled by the Lamb-Dicke parameter $\eta$.
    (c) Multimode optical cavity, where each cavity mode with profile $g_\mu(\bfr)$ dissipates at rate $\kappa$, coupled to a dissipationless driven quantum gas.
    The features of cavity dissipation are controlled by $\delta\tilde{\omega}$.
    Panels (b) and (d) show the participation ratio
    $\mathrm{PR}=(\sum_a w_a)^2/\sum_a w_a^2$,
    as functions of $\eta$ and $\delta\tilde{\omega}$, respectively, with (b) $w_a=\lambda_a\propto\lVert\hat L_a\rVert_F^2$ and
    (d) $w_\mu=\lVert\hat L_{\rm c}^{(\mu)}\rVert_F^2$.}
\end{figure}

Our main results are threefold. First, we
show that, in the dispersive regime where virtual photons mediate fermionic all-to-all interactions, atomic spontaneous emission in a single-mode cavity and photon leakage from a multimode cavity can each give rise to DQC,
providing two complementary routes to the same universal open SYK physics.
Second, the two routes start from opposite dissipationless limits: the spontaneous-emission route is integrable, whereas the multimode cavity-loss route remains chaotic~\cite{Uhrich2023}.
In the former case, quantum chaos is therefore \textit{created} by dissipation rather than merely modified by it.
Third, we trace the common emergence of DQC to a controllable increase in the number of independent dephasing channels and show that it is accompanied by experimentally accessible signatures in single-atom-resolved density dynamics.
Together, these results establish the potential of cavity QED as a hardware-efficient route to the experimental realization and control of DQC.

\paragraph{Driven-dissipative many-body cavity QED}
We consider a 2D gas of ultracold fermionic atoms confined by a trapping potential and coupled to a single optical-cavity mode, described by the bosonic annihilation (creation) operator $\hat{a}$ ($\hat{a}^\dagger$), sketched in Fig.~\ref{fig:sketch}(a).
The fermionic modes (orbitals) are inherited from the motional states of the trapping potential.
A far-detuned transverse pump illuminates the atomic cloud, driving the fermions from their ground to excited states and vice versa.
All-to-all, disordered light-matter interactions are realized by projecting a speckle potential onto the cloud~\cite{goodman_fundamental_1976}.
The Hamiltonian describing the cavity QED system reads~\cite{suppinfo}
\begin{equation}
    \hat{H} = \Delta\,\hat{a}^\dagger\hat{a} +\sum_{jk}g_{jk}(\hat{a}^\dagger + \hat{a})\hat{c}_j^\dagger\hat{c}_k, 
    \label{eq:hamiltonian}
\end{equation}
where $\Delta$ is the cavity-drive detuning and $g_{jk}$ are randomized, real couplings.
In the following, we consider the dispersive regime between the pump and the cavity mode, implying $\langle\hat{a}^\dagger\hat{a}\rangle\simeq0$, with virtual photons mediating fermionic all-to-all interactions.
The effective, fermionic-only Hamiltonian reads~\cite{mivehvar_cavity_2021,ritsch_cold_2013}
\begin{equation}
\begin{split}
    \hat{H}_{\rm eff} = -\frac{1}{\Delta}\Big[\sum_{jk}g_{jk}\hat{c}_j^\dagger\hat{c}_k\Big]^2 = -J\hat{F}^2,
    \label{eq:hamiltonian_effective}
\end{split}
\end{equation}
with $J = \Lambda^2/\Delta$ the cavity-mediated interaction scale, $\hat{F}=\sum_{jk}\widetilde{I}_{jk}\hat{c}_j^\dagger\hat{c}_k$, $g_{jk}=\Lambda\widetilde{I}_{jk}$, and $\Lambda$ the overall Raman scale.
Being proportional to the square of randomized free fermions, Eq.~\eqref{eq:hamiltonian_effective} describes an integrable system~\cite{Kim2020a}.
Several proposals to promote it to a quantum chaotic model have been put forward, from multimode cavity couplings~\cite{Uhrich2023} to time-dependent disorder~\cite{Baumgartner2024, Baumgartner2025, FerrariEtAl2026}.

Cavity QED systems, however, are also subject to dissipation~\cite{mivehvar_cavity_2021}, and must therefore be treated as open quantum systems~\cite{breuer_theory_2007}.
Dissipation is primarily governed by two processes: cavity losses at rate $\kappa$, and atomic spontaneous emission at rate $\Gamma$.
Cavity losses are described by the Lindblad jump operator $\hat{L}_{\rmc} = \sqrt{\kappa}\hat{a}$.
Since we assume $\langle\hat{a}^\dagger\hat{a}\rangle\simeq0$, this reduces to an effective atomic dephasing jump operator $\hat{L}_{\rm eff} = \sqrt{\kappa_{\rm eff}}\,\hat{F}$, where $\kappa_{\rm eff}\propto\kappa$ is an effective Purcell decay~\cite{suppinfo}.
Atomic spontaneous emission, in the regime of large atom–drive detuning (distinct from the cavity–drive detuning $\Delta$), results in elastic scattering in which a ground-state atom scatters a drive photon into the full solid angle and recoils by $k_0=2\pi/\lambda_\rmc$, where $\lambda_\rmc$ is the cavity wavelength.
The transition amplitude between different orbitals upon incoherent photon scattering is encoded by the Lamb-Dicke parameter $\eta=k_0x_0$, where $x_0$ is the typical size of the atomic orbitals.
The overall process can be modeled by a collection of dephasing jump operators $\hat{L}_\rma^{(\alpha)} = \sqrt{\Gamma_{\rm eff}}\sum_{jk}M_{jk}^{(\alpha)}\hat{c}_j^\dagger\hat{c}_k$, where $\Gamma_{\rm eff}\propto\Gamma$ and $M_{jk}^{(\alpha)}$ with $\alpha=1,\ldots,R$ form a set of randomized real dissipative couplings (correlated to the Hamiltonian overlaps $\widetilde{I}_{jk}$), whose number $R$ depends on $\eta$~\cite{suppinfo}.

Overall, the driven-dissipative dynamics of the effective fermionic system is described by the Lindblad master equation~\cite{breuer_theory_2007}
\begin{equation}
    \frac{\partial \hat{\rho}}{\partial t} = -\rmi[\hat{H}_{\rm eff}, \hat{\rho}] + \mathcal{D}[\hat{L}_{\rm eff}]\hat{\rho} + \sum_\alpha\mathcal{D}[\hat{L}_{\rm a}^{(\alpha)}]\hat{\rho}  =\mathcal{L}\hat{\rho},
    \label{eq:lindblad}
\end{equation}
with $\mathcal{D}[\hat{L}]\hat{\rho} = \hat{L}\hat{\rho}\hat{L}^\dagger - \frac{1}{2}\{\hat{L}^\dagger\hat{L},\hat{\rho}\}$ the Lindblad dissipator associated with the jump operator $\hat{L}$, $\hat{\rho}$ the system density matrix, and $\LL$ the non-Hermitian Liouvillian superoperator.
$\LL$ admits a biorthonormal eigendecomposition with right and left eigenoperators defined by $\LL\hat{\eta}_j=\lambda_j\hat{\eta}_j$ and $\LL^\dagger\hat{\sigma}_j=\lambda_j^*\hat{\sigma}_j$, and complex eigenvalues $\lambda_j$.
Our goal is to study the emergence of DQC in Eq.~\eqref{eq:lindblad}.
Notably, the closed-system limit of $\LL$ is integrable.
Therefore, if chaos arises, it must be \textit{created} by dissipation.

\paragraph{DQC and spectral form factors}
The signatures of quantum chaos are encoded in the spectral properties of the generator of the time evolution.
For Hermitian systems, the Hamiltonian eigenvalue correlations provide universally accepted signatures of chaos and integrability~\cite{dalessio_quantum_2016, haake_quantum_2018}.
These can be static (eigenvalue spacing distributions~\cite{berry_level_1977, bohigas_characterization_1984} or real spacing ratios~\cite{atas_distribution_2013} conforming to random matrix theory (RMT)) or dynamical (spectral form factors (SFFs)~\cite{leviandier_fourier_1986, wilkie_time-dependent_1991, alhassid_onset_1993, ma_correlation_1995}).
The SFF at infinite temperature can be equivalently defined as the squared modulus of the Fourier transform of the spectral density or the return probability of a coherent Gibbs state, $\textrm{SFF}(t) = |\sum_j\rme^{-\rmi E_j t}/D|^2=|\bra{\Psi}\rme^{-\rmi\hat{H}t}\ket{\Psi}|^2$, where $D$ is the Hilbert space dimension and $\ket{\Psi}=\frac{1}{\sqrt{D}}\sum_j\ket{j}$.
The SFF of a quantum chaotic system exhibits an early-time dip below its asymptotic value --- the correlation hole~\cite{ma_correlation_1995}, a dynamical signature of universal spectral rigidity --- followed by a linear ramp which converges to a plateau $\sim1/D$.
Instead, the SFF of an integrable system directly approaches $\sim1/D$.

These concepts have been generalized to open quantum systems whose dynamical generator is the non-Hermitian Liouvillian $\LL$ and spectral correlations develop among the \textit{complex} eigenvalues $\lambda_j$~\cite{grobe_quantum_1988}.
Spacing distributions and ratios have been extended to their non-Hermitian counterparts~\cite{grobe_quantum_1988, akemann_universal_2019, hamazaki_universality_2020, sa_complex_2020}; the dissipative generalization of the SFF, though, requires particular care.
In open quantum systems, universal spectral features of $\LL$ are not necessarily reflected in dynamical observables.
The dissipative $\SFF$ ($\DSFF$) introduced in Ref.~\cite{li_spectral_2021}, $\DSFF(s, t)=|\sum_j\rme^{\rmi(x_jt + y_js)}/\DD|^2$, where $\DD$ is the Liouville space dimension  and $\lambda_j = x_j + \rmi y_j$, probes the spectral correlations of $\LL$ and exhibits the correlation hole in the presence of DQC or its absence for integrable dissipative dynamics~\footnote{The DSFF introduced in Ref.~\cite{li_spectral_2021} depends on the choice of an angle in the complex plane. 
Writing $\tau=t+\rmi s$, $(t, s) = (|\tau|\cos\phi, |\tau|\sin\phi)$, at fixed $\phi$ the DSFF is the SFF of the complex eigenvalues projected along the direction $\phi$~\cite{li_spectral_2021}. 
Being 2D, it leaves freedom in the direction along which spectral rigidity is probed. 
This construction places $\Real\lambda_j$ and $\Imag\lambda_j$ on equal footing, although in the  Liouvillian evolution they determine decay rates and oscillation frequencies, respectively. 
Thus, $\phi=0$ probes decay-rate correlations, whereas $\phi=\pi/2$ probes oscillation-frequency correlations and, in the unitary limit, the Hamiltonian structure. We choose the generic direction $\phi=\pi/4$, weighting the two spectral directions equally and avoids the special symmetry axes $\phi=\{0,\pi/2\}$~\cite{li_spectral_2021}. 
For rotationally invariant spectral correlations the DSFF is independent of $\phi$, whereas anisotropies of actual Liouvillian spectra can introduce an angular dependence~\cite{garcia-garcia_universality_2023}. 
Results for the non-unfolded $\sigma$SFF and DSFF are reported in the Supplemental Material.}.
The return probability of a coherent Gibbs state has been defined for open systems~\cite{xu_thermofield_2021, matsoukas-roubeas_quantum_2024},
but does not coincide with the $\DSFF$ of Ref.~\cite{li_spectral_2021}, mainly because $\textrm{Re}\lambda_j$ enters as an exponential damping rather than a phase.
This leads to a dissipation-induced suppression of the correlation hole, regardless of the spectral structure of $\LL$.
Consequently, unlike in the Hermitian setting, universal spectral properties of the generator may remain hidden from the return dynamics.

To circumvent the difficulties posed by complex spectra, a singular value decomposition (SVD) approach to non-Hermitian systems has been considered~\cite{kawabata_singular-value_2023, nandy_probing_2025, prasad_assessment_2025}.
The SVD yields \textit{real} singular values $\sigma_j$ and orthonormal singular vectors~\cite{suppinfo}.
Ref.~\cite{roccati_diagnosing_2024} introduced the singular spectral form factor ($\sSFF$)
\begin{equation}
    \sSFF(t) = \Big|\frac{1}{\DD}\sum_j\rme^{-\rmi\sigma_j t}\Big|^2,
    \label{eq:sigma_sff}
\end{equation} to probe chaos and integrability in non-Hermitian Hamiltonians.
In a recent paper~\cite{baggioli_singular_2025}, Baggioli \textit{et al.} identified a blind spot in the SVD approach to DQC, casting doubt on the effectiveness of Eq.~\eqref{eq:sigma_sff} as a tool to unveil complex non-Hermitian quantum dynamics.

The methodological contribution of our work is to show that the $\sSFF$ can be used to diagnose the crossover between integrability and DQC in open quantum systems (where the generator is the Liouvillian and \textit{not} a non-Hermitian Hamiltonian).
In the End Matter, we further prove that at the integrable corner the blind spot of Ref.~\cite{baggioli_singular_2025} is absent, and we argue for the generality of our findings.

\begin{figure}[t]
    \centering
    \includegraphics[scale=0.5]{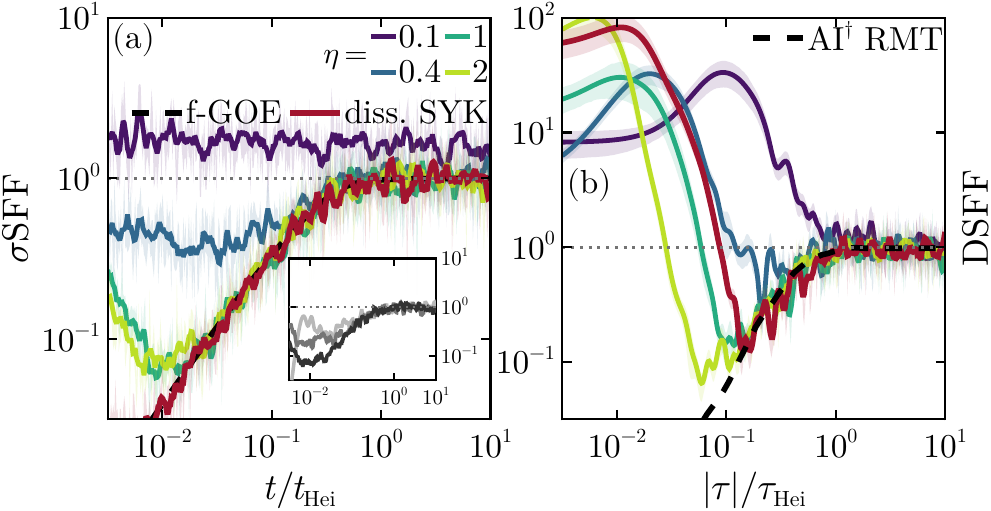}
    \caption{\label{fig:atomic_spontaneous_emission}\label{fig:single-mode-spectra}
    Single-mode spontaneous-emission route to DQC for $N=10$ and $Q=4$.
    (a) Connected unfolded $\sSFF$ at various $\eta$ (from dark blue to light green); the inset compares $N=6,8,10$, $Q=3,4,4$ (light to dark gray) at $\eta=2$.
    (b) Connected, conformally unfolded $\DSFF$ at $\phi=\pi/4$~\cite{li_spectral_2021}, with $\tau=\sqrt{t^2+s^2}\rme^{\rmi\phi}$ and $\phi=\arctan(s/t)$.
    The time axes are expressed in units of the respective Heisenberg times $t_{\rm Hei}$ and $\tau_{\rm Hei}$.
    Both form factors are normalized to their late-time plateaus $K_\infty=\mathcal{D}$.
    The red curves refer to the dissipative $\SYK$ model.
    The black reference curves are the f-GOE result in (a) and the AI$^\dagger$ RMT bulk result in (b).
    Averages are taken over 16 disorder realizations (shading represents one standard deviation).}
\end{figure}

\paragraph{DQC from atomic spontaneous emission}
We consider the open system dynamics generated by Eq.~\eqref{eq:lindblad}.
We investigate the emergence of DQC in the spectral structure of $\LL$ as the Lamb-Dicke parameter $\eta$ is varied.
Specifically, we calculate the $\sSFF$~\cite{roccati_diagnosing_2024}, the $\DSFF$~\cite{li_spectral_2021} and the complex spacing ratios (CSR)~\cite{sa_complex_2020}.
Figure~\ref{fig:atomic_spontaneous_emission}(a) reports the dynamics of the unfolded $\sSFF$ for $\eta=0.1, 0.4, 1, 2$ for $N=10$ fermionic modes and $Q=4$ particles.
At $\eta=0.1$, we are in the deep Lamb-Dicke regime; here, $\sSFF(t)\ge1$ indicates regular dissipative dynamics.
In this regime, the integrable unitary dynamics generated by Eq.~\eqref{eq:hamiltonian_effective} is subject to a single spontaneous-emission dephasing channel $\hat{L}_{\rma}^{(1)}$ [cf. Fig.~\ref{fig:sketch}(b)], in addition to the cavity-loss channel $\hat{L}_{\rm eff}$~\footnote{Notice that in Fig.~\ref{fig:atomic_spontaneous_emission}, $\mathcal{E}/\Gamma_{\rm eff}\sim\mathcal{O}(1)$, where $\mathcal{E}$ is the coherent energy scale associated with Eq.~\eqref{eq:hamiltonian_effective}. Therefore, the behavior of $\sigma$-SFF cannot be associated with a unitary integrable dynamics perturbatively disturbed by dissipation, but rather with a genuine dissipative integrable dynamics}.
We prove the integrability of this limit in the End Matter.
The situation changes qualitatively as we increase $\eta$, and, consequently, the number of spontaneous-emission jump operators $\hat{L}_{\rma}^{(\alpha)}$ [cf. Fig.~\ref{fig:sketch}(b)].
At $\eta\gtrsim1$, the $\sSFF$ develops a dip followed by a linear ramp: the universal signature of quantum chaos.
Notably, the correlation hole becomes deeper with $\eta$ (consistently with the increase in the number of spontaneous emission jump operators) and with the system size.
This last feature is reported in the inset of Fig.~\ref{fig:atomic_spontaneous_emission}(a), where we compare the dynamics of $\sSFF$ for $N=6, 8, 10$ fermionic modes and $Q=3, 4, 4$ particles, respectively.

As argued in Ref.~\cite{Baumgartner2024}, in the limit of infinite interaction rank, the integrable and dissipative parts effectively decouple. 
This motivates the comparison of the physical $\sSFF$ curves at various $\eta$ with a target dissipative SYK model with real random couplings.
Its Hamiltonian reads
\begin{equation}
    \hat{H}_4 = \sum_{jk\ell m}J_{jk\ell m}\hat{c}_j^\dagger\hat{c}_\ell^\dagger\hat{c}_k\hat{c}_m,
\end{equation}
and we consider a collection of $N$ Hermitian dephasing jump operators $\hat{L}_\alpha = \sum_{jk}K_{jk}\hat{c}_j^\dagger\hat{c}_k$.
Both $J_{jk\ell m}$ and $K_{jk}$ are sampled from zero-mean Gaussian distributions, with variances adjusted to match the physical model parameters~\cite{suppinfo}.
The $\sSFF$ associated with this target model is presented as the red curve in Fig.~\ref{fig:atomic_spontaneous_emission}(a), while the black dashed line shows the $\sSFF$ associated to random matrices drawn from the folded Gaussian Orthogonal Ensemble (f-GOE)~\cite{suppinfo}.
At $\eta\gtrsim1$, we observe that the physical model reproduces the universal behavior of the dissipative SYK and random matrix models.

The behavior of $\sSFF$ is mirrored by the $\DSFF$, plotted in Fig.~\ref{fig:atomic_spontaneous_emission}(b).
For $\eta\gtrsim1$, it develops a correlation hole, signaling rigidity in the spectrum of $\LL$.
The ramp matches both that of the dissipative SYK model (red curve) and the non-Hermitian RMT prediction for transposition-symmetric matrices, $\LL^\top=\LL$. This constraint defines the AI$^\dagger$ symmetry class and follows from the reality of the couplings~\cite{hamazaki_universality_2020, sa_symmetry_2023}.
Agreement with the universal predictions of non-Hermitian RMT is further corroborated by the CSR distribution, presented in the End Matter. 
Taken together, these results demonstrate that spontaneous emission can \textit{create} DQC in a dispersive cavity-fermion platform with controllable disorder.
Because the underlying coherent dynamics is integrable, the observed chaos is entirely dissipation-induced and it originates from the proliferation of nonlocal dephasing jump operators with increasing $\eta$.
This mechanism for the emergence of DQC was not identified in previous cavity-SYK studies, which either pursued chaos through purely Hamiltonian dynamics~\cite{Uhrich2023, Baumgartner2024} or studied how decoherence degrades an otherwise chaotic regime~\cite{FerrariEtAl2026}.

\begin{figure}[t]
    \centering
    \includegraphics[scale=0.5]{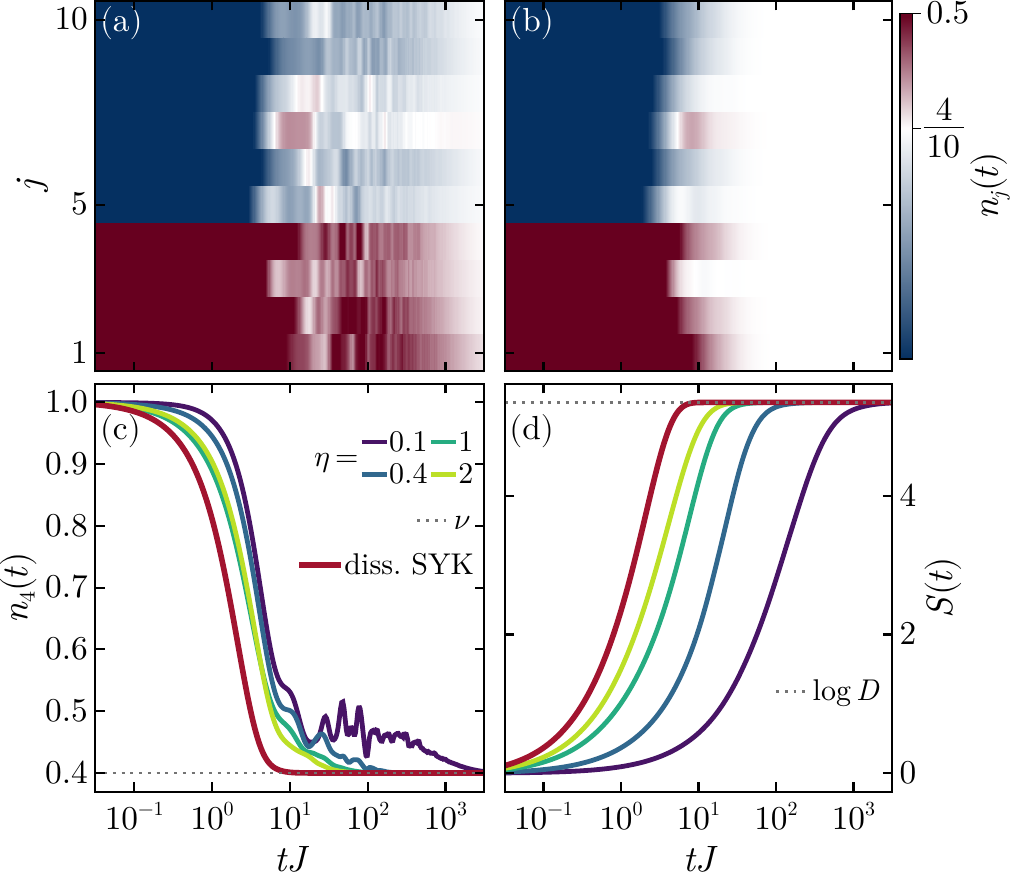}
    \caption{\label{fig:dynamics}\label{fig:single-mode-dynamics}
    Relaxation dynamics of the initial domain wall $\ket{1\dots1\,0\dots0}$ for the parameters of Fig.~\ref{fig:single-mode-spectra}.
    (a--b) Orbital occupations at (a) $\eta=0.1$ and (b) $\eta=2$.
    (c) Occupation of the initially filled orbital $j=4$ and (d) von Neumann entropy $S$ of the density matrix $\hat{\rho}(t)$.
    Curves are averaged over 16 realizations; the red curve is the dissipative $\SYK$ model.
 Dotted lines mark (c) $\nu=4/10$ and (d) $S_\infty=\log D$.}
\end{figure}

Since for open quantum systems the $\DSFF$ and the return probability of a Gibbs state are decoupled, and the $\sSFF$ is based on the auxiliary generator $\sqrt{\LL^\dagger\LL}$ rather than $\LL$, nothing guarantees that Fig.~\ref{fig:atomic_spontaneous_emission} carries information about the open-system dynamics.
It is natural to ask whether the integrable-to-chaotic crossover leaves any experimentally accessible fingerprints in the dynamics.
To answer this question, we study the dynamics of simple observables in Fig.~\ref{fig:dynamics}.
Figs.~\ref{fig:dynamics}(a-b) display the dynamics of the fermionic density $ n_j(t) = \Tr [\hat{c}_j^\dagger\hat{c}_j\,\hat{\rho}(t)]$ for $N=10$ orbitals and $Q=4$ particles, for $\eta=0.1$ and $\eta=2$, respectively.
The initial state is the domain wall $\ket{\Psi(0)}=\ket{1\dots1\,0\dots0}$.
The short- and long-time limits are independent of $\eta$: $ n_j(0) \in \{0, 1\}$ depending on $j$ for short times, and $ n_j(t\to\infty)=\nu=4/10$ (the microcanonical value at the considered filling fraction $\nu$).
This is not surprising, because dephasing eventually relaxes the system to the infinite-temperature state~\footnote{This is true everywhere except at the integrable corner discussed in the End Matter.}.
A qualitative difference in the dynamics is instead found at intermediate times. 
For small $\eta$, a prethermal plateau emerges with $n_j\ne\nu$, 
induced by the large number of conserved quantities (see the End Matter).
This dynamical regime, therefore, preserves a clear memory of the initial state.
For $\eta=2$, on the other hand, the orbital occupations rapidly relax to $\nu$ due to the fast thermalization: the growth of the dissipator rank destroys the conserved quantities characterizing the integrable corner.
The contrast between the two regimes persists over a broad temporal window.
The smooth disappearance of the prethermal phase with increasing $\eta$ and the approach to the fast thermalization dynamics of the target model are shown in Fig.~\ref{fig:dynamics}(c). 
Here, we focus on a single orbital ($j=4$) and plot its occupation $n_4(t)$ as a function of time for different $\eta$; the red curve is the reference dissipative SYK.
This $\eta$-controlled crossover in the dynamics of a simple observable can be an experimental probe of increasing dissipative rank and loss of conserved quantities, as it relies on single-atom resolution techniques~\cite{holten_observation_2021, holten_observation_2022}.
However, it does not constitute a strict signature of spectral correlations or chaos.
Finally, in Fig.~\ref{fig:dynamics}(d) we plot the von Neumann entropy of the density matrix, $S(t) = -\Tr [\hat{\rho}(t)\log\hat{\rho}(t)]$, which measures the mixedness of $\hat{\rho}(t)$, hence quantifying the distance from the closed-system limit ($S(0)=0$) and the maximally mixed infinite-temperature state ($S(t\to\infty)=\log D$).
$S(t)$ converges to $\log D$ as soon as $ n_4(t)\to\nu$, with $S(t)$ slowly growing in the prethermal phase.

\paragraph{DQC from cavity dissipation}
We now discuss an alternative route to DQC based solely on cavity losses.
To this end, we consider the driven-dissipative multimode optical cavity sketched in Fig.~\ref{fig:sketch}(c).
Such platforms are experimentally accessible and have already been shown to capture many-body phenomena~\cite{Kroeze2023, Kroeze2025, Marsh2025}.
Multimode cavities have previously been proposed as platforms for unitary SYK physics~\cite{Uhrich2023}.
In this setting, Eq.~\eqref{eq:hamiltonian} is replaced by
\begin{equation}
    \hat{H} = \sum_{\mu=1}^M\Big[\Delta_\mu\,\hat{a}_\mu^\dagger\hat{a}_\mu +\sum_{jk}g_{jk}^{(\mu)}(\hat{a}_\mu^\dagger + \hat{a}_\mu)\hat{c}_j^\dagger\hat{c}_k\Big], 
    \label{eq:hamiltonian_multimode}
\end{equation}
where $\hat{a}_\mu$ ($\hat{a}_\mu^\dagger$) is the annihilation (creation) operator for the $\mu$-th cavity mode.
In the dispersive regime the effective fermionic-only Hamiltonian approximates the chaotic SYK Hamiltonian, $\hat{H}_{\rm eff} =\sum_{jk\ell m}g_{j\ell km}\hat{c}_j^\dagger\hat{c}_\ell^\dagger\hat{c}_k\hat{c}_m$, with $g_{j\ell km} = \frac12\sum_{\mu}\big(g_{jk}^{(\mu)}g_{\ell m}^{(\mu)} - g_{jm}^{(\mu)}g_{\ell k}^{(\mu)}\big)/\Delta_\mu$~\cite{Uhrich2023}.
We consider here cavity losses, modeled by $\hat{L}_\mu = \sqrt{\kappa}\hat{a}_\mu$, to be the sole source of decoherence.
In the dispersive regime, they effectively map into a collection of dephasing jump operators $\hat{L}_{\rm eff}^{(\mu)} = \sqrt{\kappa_{\rm eff}}\sum_{jk}\widetilde{I}_{jk}^{(\mu)}\hat{c}_j^\dagger\hat{c}_k$, where $\kappa_{\rm eff}\propto\kappa$.
The fundamental difference from the single-mode setup considered above is that now the dissipationless limit can be maximally chaotic.
The control parameter is $\delta\tilde{\omega} = \delta\omega/\Delta$, with $\delta\omega$ the frequency difference between two successive cavity modes (assumed to be independent of the mode index $\mu$).
$\delta\tilde{\omega}$ effectively controls the number $M$ of cavity modes participating in the dynamics [cf. Fig.~\ref{fig:sketch}(d)], and therefore the number of effective jump operators originating from cavity losses.

\begin{figure}[t]
    \centering
    \includegraphics[scale=0.5]{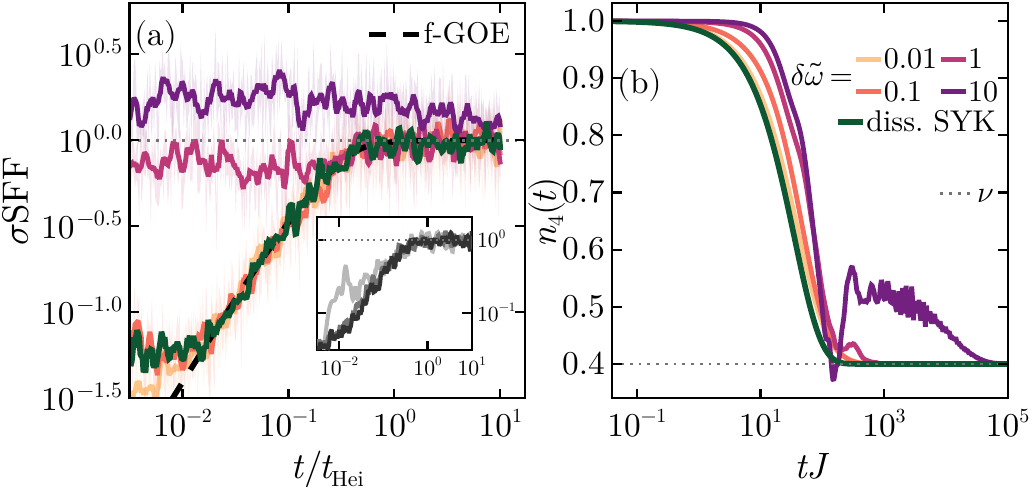}
    \caption{\label{fig:cavity_dissipation}\label{fig:multimode-sigma-sff}
    Multimode cavity-loss route to DQC for $N=10$ and $Q=4$.
    (a) Connected unfolded $\sSFF$ at various $\delta\tilde{\omega}$ (from light pink to dark purple), normalized by $K_\infty=D$ and with time expressed in units of $t_{\rm Hei}$, for $\delta\tilde{\omega}=0.01,0.1,1,10$.
    The inset compares $N=6,8,10$, $Q=3,4,4$ (light to dark gray) at $\delta\tilde{\omega}=0.01$.
    (b) Occupation $n_4(t)$ of the initially filled orbital; the dotted line marks $\nu=4/10$.
    All parameters and conventions as in Fig.~\ref{fig:atomic_spontaneous_emission}.
    Averages are taken over 16 disorder realizations (shading represents one standard deviation).}
\end{figure}

In Fig.~\ref{fig:cavity_dissipation}(a) we plot the unfolded $\sSFF$ with increasing $\delta\tilde{\omega}$ for $N=10$ and $Q=4$. At $\delta\tilde{\omega}=10$, the system is effectively in the single-mode limit: $\sSFF(t)\ge 1$, consistently with integrable dynamics obtained from Eq.~\eqref{eq:lindblad} without the spontaneous emission term. As we decrease $\delta\tilde{\omega}$, more cavity modes are involved in the driven-dissipative dynamics and a correlation hole opens in $\sSFF(t)$.
At $\delta\tilde{\omega}=0.01$, the $\sSFF$ is indistinguishable from the f-GOE (black dashed curve) and dissipative $\SYK$ (green curve) references.
The inset of Fig.~\ref{fig:cavity_dissipation}(a) shows the increasing depth of the correlation hole with the system size. 
This crossover towards universal non-Hermitian RMT with decreasing $\delta\tilde{\omega}$, driven by cavity dissipation, is also captured by the CSR distribution reported in the End Matter, consistent with the same non-Hermitian universality class induced by atomic spontaneous emission.
This is another central finding of our work: despite starting from opposite closed-system limits (the integrable Eq.~\eqref{eq:hamiltonian_effective} and the SYK model~\cite{Uhrich2023}), the two routes, controlled respectively by spontaneous emission (via $\eta$) and multimode cavity loss (via $\delta\tilde{\omega}$), converge to the same universal non-Hermitian RMT behavior~\footnote{They are not, however, completely equivalent. 
Tuning $\eta$ increases the number of jump operators leaving $\hat{H}_{\rm eff}$ unchanged. 
Tuning $\delta\tilde{\omega}$ changes \textit{both} the dissipator and the Hamiltonian.}.
The rank of the Lindblad dissipator [cf. Figs.~\ref{fig:sketch}(b,d)] is the microscopic mechanism responsible for the crossover, and the details of the setup do not alter the emerging universal features. 

Finally, in Fig.~\ref{fig:cavity_dissipation}(b) we investigate whether the integrable-to-chaotic crossover driven by cavity losses leaves fingerprints in the dissipative dynamics.
Adopting the same protocol used in Fig.~\ref{fig:dynamics}, we show it does. At $\delta\tilde{\omega}=10$ an extended prethermal phase retains memory of the initial domain wall, which is suppressed by decreasing $\delta\tilde{\omega}$: at $\delta\tilde{\omega}=0.01$ the fermionic density rapidly relaxes to the microcanonical value $\nu$, overlapping the dissipative SYK reference (green curve), in agreement with the onset of spectral universality in $\LL$.
This suggests that the complementary route towards DQC discussed above leaves experimentally accessible traces.

\paragraph{Discussion}
We have shown that dissipative SYK universality can emerge in cavity QED platforms by harnessing two native dissipative mechanisms: atomic spontaneous emission in a single-mode cavity and photon leakage in a multimode cavity.
The two complementary routes start from opposite closed-system limits (integrable for the single-mode setup and chaotic for the multimode one) yet converge to the same universal non-Hermitian RMT behavior, as captured by several DQC diagnostics.
This is particularly remarkable for the single-mode setup, where DQC arises entirely from dissipation.
The integrability-to-chaos crossover is accompanied by an experimentally accessible fingerprint in single-atom-resolved density measurements: an extended prethermal regime gives way to fast thermalization.

Importantly, our proposal circumvents two major challenges identified in previous work~\cite{Baumgartner2024, FerrariEtAl2026}: no complex time-dependent protocol is needed to generate Hamiltonian chaos, and dissipation, rather than a threat to fragile quantum states, is the resource that creates chaotic behavior. Further challenges arising in realistic setups, together with possible workarounds, are discussed in the Supplemental Material.

Finally, our results suggest a connection to holography. The low-energy
dynamics of the SYK model is described by nearly-AdS$_2$
gravity~\cite{maldacena_remarks_2016, rosenhaus_introduction_2019}, and
the ramp of the SFF is reproduced on the gravity side by wormhole
saddles~\cite{saad2019semiclassicalrampsykgravity}, which in the
dissipative SYK model become Keldysh
wormholes~\cite{garcia-garcia_keldysh_2023}. Cavity QED devices, where
the dissipative rank can be tuned \textit{in situ}, could therefore provide a
controllable setting to explore such open-system gravitational saddles.

\paragraph{Acknowledgements}
We acknowledge useful discussions with Marco Schirò, Gaia Stella Bolognini, Pietro Pelliconi, Bastien Lapierre, Andrea Legramandi and Jean-Philippe Brantut.
P.P., F.F., M.S.\@ and V.S.\@ acknowledge support from the Swiss National Science Foundation through Projects Nos.\@ 200020\_215172, 200021-227992, and 20QU-1\_215928, and as part of NCCR SPIN (grant number 225153).
M.S.\@ acknowledges funding from the Swiss Academy of Sciences (SCNAT) through the Swiss Quantum Initiative (SQI) Grant No.\@ 24\_1111.
Large language models (LLMs), including GPT-6 Astra (OpenAI) and Claude Opus 5.5 (Anthropic), were employed to assist with code development, figure preparation, and proofreading of the manuscript. The original ideas and the mathematical derivations behind the results are the authors' own. All the material was reviewed by the authors, who take full responsibility for the content of this work.

% Following: https://journals.aps.org/authors/data-availability-statements
\paragraph{Data availability}
The data that support the findings of this study were generated by numerical simulations. The source code and parameters used to generate the simulations, as well as the data used to generate the figures, are publicly available~\cite{dissipativesyk-dataset}.

\paragraph{Note added}
While finalizing this manuscript, Ref.~\cite{pelliconi2026dissipationenhancedscramblingsykmodel} was announced on arXiv. 
It shows that bosonic loss in the Yukawa-SYK model preserves a positive Lyapunov exponent and can enhance the scrambling rate. That work focuses on operator scrambling in a distinct model, whereas here we characterize the Liouvillian spectral structure and show that dissipation can create chaos from an integrable closed-system limit.

\bibliography{references}

\clearpage

\onecolumngrid
\begin{center}
  \textbf{\large End Matter}\\[1em]
\end{center}
\vspace{1em}
\twocolumngrid

\paragraph{Analytical results for the integrable corners}
Both the spontaneous emission and cavity dissipation routes possess integrable dissipative corners at $\eta\to0$ and $\delta\tilde{\omega}\to\infty$, respectively.
Here we show that these corners are amenable to analytical calculations and immune from the blind spot for the $\sSFF$ identified in Ref.~\cite{baggioli_singular_2025}.

\begin{figure}[t!]
 \centering
 \includegraphics[width=\columnwidth]{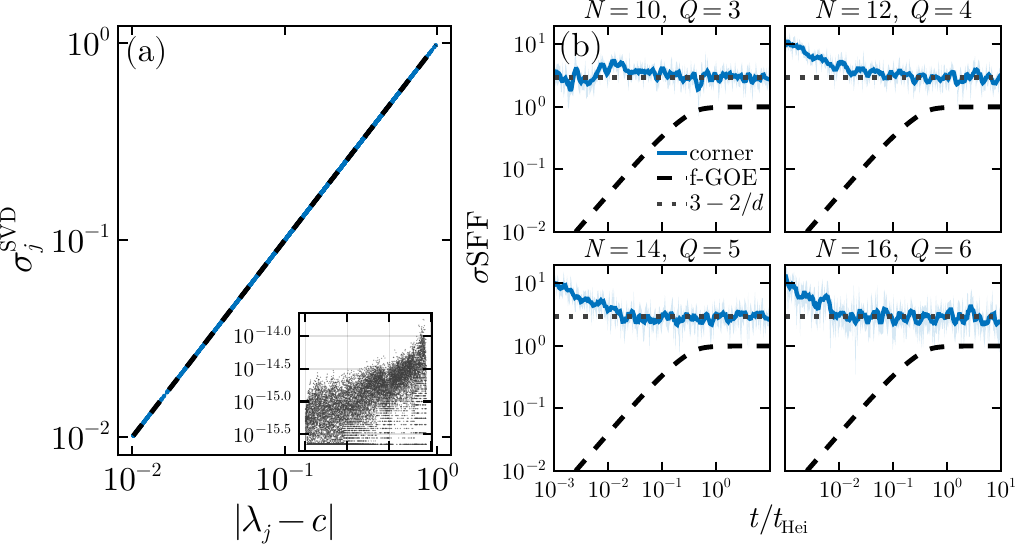}
 \caption{\label{fig:integrable_corner}
 Integrable corner at $\Gamma_{\rm eff}/J=1$ and $\kappa_{\rm eff}/J=0.2$.
 (a) Singular values of $\overline{\mathcal{L}}$ at $N=10$, $Q=4$ versus the analytical prediction $\lvert\lambda_j-c\rvert$ (blue), black dashed $45^\circ$ reference line; the inset shows $\lvert\sigma_j^\mathrm{SVD}-\lvert\lambda_j-c\rvert\rvert$.
 (b) Connected unfolded $\sSFF$ $(N,Q)=(10,3),(12,4),(14,5),(16,6)$ (blue), with the analytical f-GOE reference (dashed) and the predicted plateau $3-2/D$ (dotted).
 Curves are averaged over $16,16,12,8$ realizations, respectively; shading denotes one standard deviation.}
\end{figure}

In both cases, the coherent part of the system coincides with Eq.~\eqref{eq:hamiltonian_effective}: the square of a quadratic Hamiltonian, $\hat{H}_{\rm eff}=-J\hat{F}^2$, $\hat{F} = \sum_{jk}\widetilde{I}_{jk}\hat{c}_j^\dagger\hat{c}_k$.
Diagonalizing the coupling matrix $\widetilde{I}$ and introducing the corresponding fermionic modes $\hat{d}_a$, we obtain $\hat{F} = \sum_a\varepsilon_a\hat{n}_a$, $\hat{n}_a = \hat{d}_a^\dagger\hat{d}_a$, which is diagonal in the occupation basis $\{\ket{m}\}$ of the eigenmodes.
Restricting to the $Q$-particle sector, and labeling the Fock states by their occupation vectors $n^{(m)} = \big(n_1^{(m)},\ldots,n_N^{(m)}\big)$, the Hamiltonian spectrum reads $E_m = -Jf_m^2$, $f_m = \sum_a\varepsilon_an_a^{(m)}$, $m=1,\ldots,D$, with $D=\binom{N}{Q}$.
The fermionic jump operator corresponding to cavity loss is $\hat{L}_{\rm eff} = \sqrt{\kappa_{\rm eff}}\,\hat{F}$, while atomic spontaneous emission in the $\eta\to0$ limit yields the single jump operator $\hat{L}_{\rm a}^{(1)} = \sqrt{\Gamma_{\rm eff}}\,\hat{F}$~\cite{suppinfo}. 
Therefore, $\mathcal{D}[\hat{L}_{\rm eff}] + \mathcal{D}[\hat{L}_{\rm a}^{(1)}] = \mathcal{D}[\sqrt{\Gamma_{\rm tot}}\hat{F}]$ with $\Gamma_{\rm tot} = \kappa_{\rm eff} + \Gamma_{\rm eff}$.
Henceforth, we measure all frequencies and decay rates in units of $J$, and time in units of $J^{-1}$; we therefore set $J=1$ and use the same symbols for the resulting dimensionless quantities.
The vectorized Liouvillian acquires the compact form $\LL = \rmi\HH - \frac{\Gamma_{\rm tot}}{2}\JJ^2$, $\HH = \hat{H}\otimes\mathds{1} - \mathds{1}\otimes\hat{H}$, $\JJ = \hat{F}\otimes\mathds{1} - \mathds{1}\otimes\hat{F}$.
Notice that here Jordan-Wigner signs arising from the vectorization of fermionic superoperators~\cite{kawabata_symmetry_2023} are absent because $\hat{H}$ and all jump operators always involve an even number of fermionic operators.
$\LL$ is normal, i.e., $[\LL, \LL^\dagger]=0$, because $[\JJ^2,\HH]=0$.
In this model, non-normality of $\LL$ is driven by the proliferation of the spontaneous-emission jump operators with $\eta$.
The right eigenstates of $\LL$ are the operators $\ket{m}\bra{n}$, $m,n=1,\ldots,D$ and the exact eigenvalues read
\begin{equation}
    \lambda_{mn} = \rmi J(f_m^2-f_n^2)-\frac{\Gamma_{\rm tot}}{2}(f_m-f_n)^2.
    \label{eq:integrable_eigenvalues}
\end{equation}
The trace centering term, necessary for SVD calculations~\cite{suppinfo}, is given analytically by:
\begin{equation}
    c = \frac{\Tr \LL}{\DD} = -\Gamma_{\rm tot}\textrm{var}(f), 
\end{equation}
where $\textrm{var}(f)=\sum_m\frac{f_m^2}{D} - \sum_{mn}\frac{f_mf_n}{D^2}$.
We can now calculate the singular values. 
A normal matrix is unitarily diagonalizable, hence, having defined the trace-centered Liouvillian $\overline{\LL} = \LL - c\mathds{1}$, we get $\overline{\LL} = \UU\,\textrm{diag}(\lambda_{mn}-c)\,\UU^\dagger$ and $\overline{\LL}^\dagger\overline{\LL} = \UU\,\textrm{diag}(|\lambda_{mn}-c|^2)\,\UU^\dagger$. 
Therefore
\begin{equation}
    \sigma_{mn} = |\lambda_{mn}-c| = \sqrt{\left[\frac{\Gamma_{\rm tot}}{2}u_{mn}^2 - \Gamma_{\rm tot}\textrm{var}(f)\right]^2+J^2u_{mn}^2s_{mn}^2},
    \label{eq:integrable_singular_values} 
\end{equation}
where $u_{mn} = f_m-f_n$ and $s_{mn} = f_m+f_n$.
This analytical result is verified in Fig.~\ref{fig:integrable_corner}(a).

\begin{figure*}[t!]
\centering
\includegraphics[width=\textwidth]
{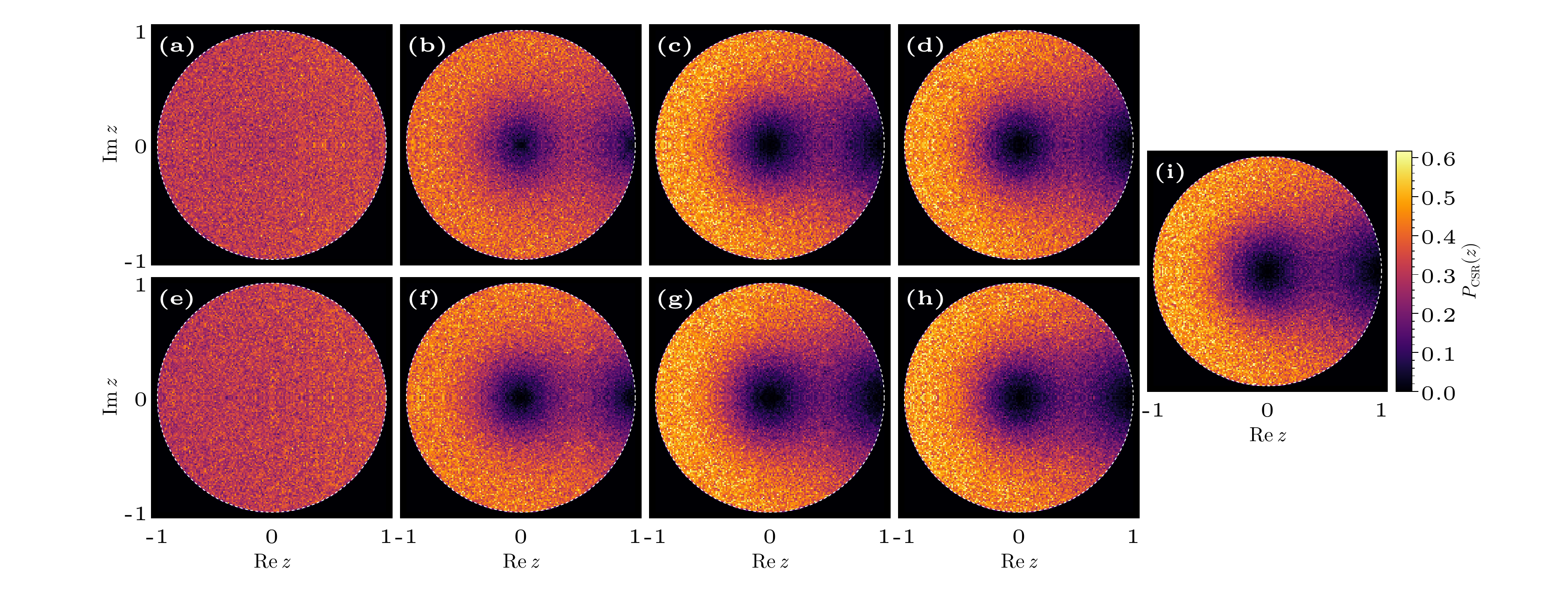}
\caption{\label{fig:complex_spacing_ratios}
CSR distributions for $N=10$, $Q=4$, and $16$ disorder realizations.
Panels (a-d) show the single-mode spontaneous-emission Liouvillian at $\eta=0.1$, $0.4$, $1$, $2$ respectively; panels (e-h) show the multimode cavity-loss Liouvillian at $\delta\tilde{\omega}=10$, $1$, $0.1$, $0.01$ respectively; panel (i) shows the target dissipative SYK model.
The steady-state eigenvalue with $|\lambda|\leq10^{-8}$ and the numerically real BDI$^\dagger$ subset with $|\operatorname{Im}\lambda|<10^{-6}$ are removed and ratios are evaluated only for reference eigenvalues in the bulk region $|\lambda-c|\leq\tfrac{1}{2}r_{\max}$ around the spectral centroid $c$. 
Color denotes the normalized spacing-ratio density $P_{\rm CSR}(z)$; the outer dashed circle marks $|z|=1$.
The moments $\langle r \rangle$ and $-\langle\cos\theta\rangle$ are collected in Table~\ref{tab:csr-moments}.}
\end{figure*}

\begin{table*}[ht]
    \centering
    \caption{Complex-spacing-ratio moments for Fig.~\ref{fig:complex_spacing_ratios}.}
    \label{tab:csr-moments}
    \setlength{\tabcolsep}{7pt}
    \begin{tabular}{@{}l *{9}{c}@{}}
        \toprule
        & \multicolumn{4}{c}{Atomic spontaneous emission}
        & \multicolumn{4}{c}{Multimode cavity loss}
        & \multicolumn{1}{c}{Reference} \\
        \cmidrule(lr){2-5}
        \cmidrule(lr){6-9}
        \cmidrule(l){10-10}
        Moment
        & $\eta=0.1$
        & $\eta=0.4$
        & $\eta=1$
        & $\eta=2$
        & $\delta\tilde{\omega}=10$
        & $\delta\tilde{\omega}=1$
        & $\delta\tilde{\omega}=0.1$
        & $\delta\tilde{\omega}=0.01$
        & Dissipative SYK \\
        \midrule
        $\langle r\rangle$
        & $0.670$ & $0.703$ & $0.721$ & $0.720$
        & $0.668$ & $0.713$ & $0.722$ & $0.721$
        & $0.722$ \\
        $-\langle\cos\theta\rangle$
        & $-0.018$ & $0.105$ & $0.183$ & $0.181$
        & $-0.024$ & $0.120$ & $0.192$ & $0.194$
        & $0.191$ \\
        \bottomrule
    \end{tabular}
\end{table*}

Equations \eqref{eq:integrable_eigenvalues} and \eqref{eq:integrable_singular_values} yield several consequences.
First, $\lambda_{mm}=0$: the steady-state manifold is $D$-fold degenerate.
In addition, $\lambda_{mn} = \lambda_{nm}^*$ and $\sigma_{mn} = \sigma_{nm}$ (every off-diagonal singular value is at least 2-fold degenerate).
These two properties have an immediate implication on $\sSFF$. 
At late times, $\sSFF(t) = \Big|\sum_j\rme^{-\rmi\sigma_jt}\Big|^2=\sum_jb_j^2$, where $b_j$ are the singular-value multiplicities.
The number of singular values is, by definition, $\sum_jb_j$.
Now, the $D(D-1)$ off-diagonal values form $D(D-1)/2$ groups of multiplicity 2, whereas the $D$ diagonal values form one group of multiplicity $D$.
Hence, $\sum_jb_j^2=2D(D-1) + D^2=3D^2-2D$ and $\sum_jb_j=D(D-1)+D=D^2$.
The late-time plateau of the connected $\sSFF$ is therefore $K_\infty = \sum_jb_j^2/\sum_jb_j = 3-\frac{2}{D}$ and \textit{not} 1.
The analytical plateau and the absence of the ramp for the integrable corner are tested in Fig.~\ref{fig:integrable_corner}(b) against several system sizes.
This finding explains the behavior of $\sSFF$ in Figs.~\ref{fig:atomic_spontaneous_emission}(a) and \ref{fig:cavity_dissipation}(a) for small $\eta$ and large $\delta\tilde{\omega}$, respectively.

Finally, we can explain why our integrable corner is immune from the blind spot discussed in Ref.~\cite{baggioli_singular_2025}.
The mechanism behind the blind spot is that Hermitianization is a nonlinear operation.
For non-Hermitian random free fermions, the SVD manufactures an interacting operator out of a free one~\cite{baggioli_singular_2025}.
At our integrable corner, Hermitianization acts on the spectrum as the pointwise map $\lambda\mapsto|\lambda-c|$ and no eigenvalue mixing (responsible for the destruction of the Poissonian statistics) occurs, so the failure mode of Ref.~\cite{baggioli_singular_2025} cannot arise.
We therefore conjecture that the SVD approach remains faithful when the non-Hermitian generator is normal in the integrable limit.

\paragraph{Complex spacing ratios}
The standard signature of DQC are the CSRs, introduced in Ref.~\cite{sa_complex_2020} and defined as
\begin{equation}
    z_j = \frac{\lambda_j-\lambda_j^{\rm NN}}{\lambda_j-\lambda_j^{\rm NNN}} = r_j\,\rme^{\rmi\theta_j},
\end{equation}
where $\lambda_j^{\rm NN}$ ($\lambda_j^{\rm NNN}$) is the nearest neighbor (next-to-nearest neighbor) of $\lambda_j$.
The spectral bulk of the spontaneous-emission Liouvillian, due to transposition symmetry $\LL^\top=\LL$ imposed by the real Hermitian couplings, belongs to the AI$^\dagger$ symmetry class of non-Hermitian random matrices~\cite{hamazaki_universality_2020, suppinfo, sa_symmetry_2023}.
The CSR distribution can resolve universality classes~\cite{sa_complex_2020}, especially via the single-number indicators $\langle r \rangle$ and $-\langle\cos\theta\rangle$.
For integrable dissipative systems $\langle r \rangle=2/3$ and $-\langle\cos\theta\rangle=0$, while for non-Hermitian random matrices in the AI$^\dagger$ class $\langle r \rangle=0.722$ and $-\langle\cos\theta\rangle=0.195$~\cite{AkemannEtAl2026CSR}.

Figures~\ref{fig:complex_spacing_ratios}(a-d) show the CSR distributions of the Liouvillian corresponding to the single-mode setup accounting for spontaneous emission at $\eta=0.1$, $0.4$, $1$, $2$, respectively.
As $\eta$ increases, the distribution crosses from an almost flat profile to the characteristic non-Hermitian random-matrix form, whose depleted regions reflect angular eigenvalue repulsion.
Figs.~\ref{fig:complex_spacing_ratios}(e-h) display the CSR distributions for the multimode cavity-loss Liouvillian at $\delta\tilde{\omega}=10$, $1$, $0.1$, $0.01$, respectively.
Decreasing $\delta\tilde{\omega}$ produces the same crossover toward the universal AI$^\dagger$ bulk profile.
Fig.~\ref{fig:complex_spacing_ratios}(i) shows the symmetry-matched target model, i.e., the real dissipative SYK model, which lies close to the chaotic single- and multimode ensembles.
Table~\ref{tab:csr-moments} quantifies these trends.

\clearpage
\onecolumngrid
\setcounter{page}{1}
\thispagestyle{empty}
\setcounter{equation}{0}
\setcounter{figure}{0}
\setcounter{table}{0}
\setcounter{section}{0}
\renewcommand{\thesection}{S.\arabic{section}}
\renewcommand{\thesubsection}{\thesection.\arabic{subsection}}
\renewcommand{\thesubsubsection}{\thesubsection.\arabic{subsubsection}}
\renewcommand{\theequation}{S\arabic{equation}}
\renewcommand{\thefigure}{S\arabic{figure}}
\renewcommand{\thetable}{S\arabic{table}}
\renewcommand{\theHequation}{S\arabic{equation}}
\renewcommand{\theHfigure}{S\arabic{figure}}
\renewcommand{\theHtable}{S\arabic{table}}
\titleformat{\section}{\bfseries\centering\MakeUppercase}{\thesection.}{1em}{}
\titleformat{\subsection}{\bfseries\centering}{\thesubsection}{1em}{}
\titleformat{\subsubsection}{\centering\itshape}{\thesubsubsection}{1em}{}
\makeatletter
\renewcommand{\p@subsection}{}
\renewcommand{\p@subsubsection}{}
\makeatother
\setcounter{secnumdepth}{3}

\begin{center}
  {\large\bfseries Supplemental Material for \\ ``Dissipation-induced Sachdev-Ye-Kitaev physics in many-body \\ cavity quantum electrodynamics''\par}
  \vspace{1.15em}
  {\normalsize
  Pietro Pacchioni\,\orcidlink{0000-0001-7496-3836},
  Filippo Ferrari\,\orcidlink{0009-0003-6317-0816},
  Vincenzo Savona\,\orcidlink{0000-0002-8984-6584},
  and Matteo Seclì\,\orcidlink{0000-0002-9608-096X}\par}
  {\small\itshape Institute of Physics and Center for Quantum Science and Engineering,\\ École Polytechnique Fédérale de Lausanne (EPFL), Lausanne, Switzerland\par}
\end{center}
\vspace{1em}

\section{Derivation of the effective models}
\label{app:effective_models}

\subsection*{A. Single-mode cavity}
\label{sec:single-mode-cavity}

\subsubsection*{1. Hamiltonian}
\label{sec:single-mode-hamiltonian}

We consider fermionic atoms confined by a 2D harmonic potential and coupled to a single optical-cavity mode. In the frame rotating at the drive frequency $\omega_\rmd$, the microscopic Hamiltonian can be written as ($=1$)
\begin{equation}
\begin{split}
\hat H={}
\Delta\,\hat a^\dagger\hat a
+\sum_{s={\rm g,e}}\int \rmd^2r\,
\hat{\Psi}_{s}^{\dagger}(\bfr)
\hat h_{\rm t}
\hat{\Psi}_{s}(\bfr)
-\int \rmd^2r\,\Delta_{\rm da}(\bfr)
\hat{\Psi}_{\rm e}^{\dagger}(\bfr)
\hat{\Psi}_{\rm e}(\bfr)
+\int \rmd^2r\,
\left[
\hat{\Phi}(\bfr)
\hat{\Psi}_{\rm e}^{\dagger}(\bfr)
\hat{\Psi}_{\rm g}(\bfr)
+\hat{\Phi}^{\dagger}(\bfr)
\hat{\Psi}_{\rm g}^{\dagger}(\bfr)
\hat{\Psi}_{\rm e}(\bfr)
\right],
\label{eq:supp_rotating_hamiltonian}
\end{split}
\end{equation}
where
\begin{equation}
\hat h_{\rm t}
=
-\frac{\nabla^2}{2m_{\rm at}}
+\frac{m_{\rm at}\omega_{\rm t}^2|\bfr|^2}{2},
\qquad\qquad
\hat{\Phi}(\bfr)
=
\Omega_\rmd g_\rmd(\bfr)
+\frac{\Omega}{2}g_{\rm c}(\bfr)\hat a\,,
\end{equation}
and $\Delta=\omega_\rmc-\omega_\rmd$.
Here $g_{\rm d}(\bfr)$ and $g_{\rm c}(\bfr)$ are the drive- and cavity-mode profiles, respectively, while $\omega_{\rm t}$ and $m_{\rm at}$ are the trapping frequency and the atomic mass, respectively. The spatially dependent atom-drive detuning is generated by an AC-Stark speckle field as detailed in Ref.~\cite{FerrariEtAl2026}.
For $|\Delta_{\rm da}(\boldsymbol r)|$ much larger than the remaining microscopic energy scales, the atomic excited states follow the ground states adiabatically:
\begin{equation}
\hat{\Psi}_{\rm e}(\bfr)
\simeq
\frac{\hat{\Phi}(\bfr)}
{\Delta_{\rm da}(\bfr)}
\hat{\Psi}_{\rm g}(\bfr).
\label{eq:supp_excited_elimination}
\end{equation}
Substituting Eq.~\eqref{eq:supp_excited_elimination} into Eq.~\eqref{eq:supp_rotating_hamiltonian} yields
\begin{equation}
\hat H
=
\Delta\hat a^\dagger\hat a
+\int \rmd^2r\,
\hat{\Psi}_{\rm g}^{\dagger}(\bfr)
\hat h_{\rm t}
\hat{\Psi}_{\rm g}(\bfr)
+
\int \rmd^2r\,
\frac{
\hat{\Phi}^{\dagger}(\bfr)
\hat{\Phi}(\bfr)}
{\Delta_{\rm da}(\bfr)}
\hat{\Psi}_{\rm g}^{\dagger}(\bfr)
\hat{\Psi}_{\rm g}(\bfr).
\label{eq:supp_atom_eliminated_hamiltonian}
\end{equation}
The last term contains a pump-induced one-body potential, a photon-number-dependent dispersive shift, and the pump--cavity Raman process. 
The pump-induced potential is compensated as detailed in Ref.~\cite{Uhrich2023}, while the photon-number-dependent contribution is neglected for
$\langle\hat a^\dagger\hat a\rangle\simeq0$.
Introducing the oscillator length
$x_0=(m_{\rm at}\omega_{\rm t})^{-1/2}$ and the dimensionless coordinate
$\bfu=\bfr/x_0$, we expand the fermionic field operator in the basis of single-particle wave functions
\begin{equation}
\hat{\Psi}_{\rm g}(\bfr)
=
\frac{1}{x_0}
\sum_{j=1}^{N}
\phi_j(\bfu)\hat c_j.
\end{equation}
The orbitals are products of one-dimensional harmonic-oscillator eigenfunctions,
\begin{equation}
\phi_j(\bfu)
=
\psi_{n_x^{(j)}}(u_x)
\psi_{n_y^{(j)}}(u_y),
\qquad\qquad
\psi_n(u)
=
\frac{
\rme^{-u^2/2}H_n(u)}
{\pi^{1/4}\sqrt{2^n n!}}\,.
\label{eq:supp_ho_wavefunction}
\end{equation}
The many-body physics is encoded in the anticommuting fermionic annihilation (creation) operators $\hat{c}_j$ ($\hat{c}_j^\dagger$).
The fermion number
$\hat Q=\sum_j\hat c_j^\dagger\hat c_j$
is conserved, and the dynamics can therefore be restricted to a fixed-$Q$ sector of dimension
$D=\binom{N}{Q}$.
It is useful to separate the overall Raman scale from the dimensionless spatial overlaps:
\begin{equation}
\Lambda
=
\frac{\Omega\Omega_{\rm d}}{2\Delta_{\rm da}},
\qquad\qquad
\widetilde I_{jk}
=
\int \rmd^2u\,
w(\bfu)
\phi_j^*(\bfu)
\phi_k(\bfu),
\qquad\qquad
g_{jk}
=
\Lambda\widetilde I_{jk},
\label{eq:supp_single_mode_overlap}
\end{equation}
where
\begin{equation}
w(\bfu)
=
\frac{
g_{\rm d}(\bfu)g_{\rm c}(\bfu)}
{\Delta_{\rm da}(\bfu)/\Delta_{\rm da}}
\end{equation}
contains the spatially disordered detuning and the optical-mode profiles. In the numerical calculations the drive and cavity waists are much larger than the atomic cloud, so that
$g_{\rm d}(\bfu)\simeq g_{\rm c}(\bfu)\simeq1$.
For real mode profiles and harmonic-oscillator orbitals, the couplings $g_{jk}$ are real.
We also neglect the (trivially integrable) trap Hamiltonian coming from $\hat{h}_{\rm t}$ (see the discussion below).
In this paper, we use harmonic oscillator wave functions, but we expect our results to be largely independent of the chosen single-particle basis. 
After these approximations, the remaining second quantized Hamiltonian is
\begin{equation}
\hat H
=
\Delta\,\hat a^\dagger\hat a
+
\Lambda(\hat a+\hat a^\dagger)\hat F,
\qquad\qquad
\hat F
=
\sum_{jk}\widetilde I_{jk}\hat c_j^\dagger\hat c_k,
\label{eq:supp_single_mode_hamiltonian}
\end{equation}
which is Eq.~\eqref{eq:hamiltonian}. For
$|\Delta|\gg|g_{jk}|$, the cavity field can be eliminated via a Schrieffer-Wolff transformation, which gives
\begin{equation}
\hat H_{\rm eff}
=
-J\hat F^2
=
-J
\Big[
\sum_{jk}
\widetilde I_{jk}
\hat c_j^\dagger\hat c_k
\Big]^2,
\qquad\qquad
J
=
\frac{\Lambda^2}{\Delta},
\label{eq:single-mode-square}
\end{equation}
Normal ordering makes the two-body and one-body contributions explicit:
\begin{equation}
\begin{split}
-J\hat F^2
=
J
\sum_{jk\ell m}
\widetilde I_{jk}\widetilde I_{\ell m}
\hat c_j^\dagger
\hat c_\ell^\dagger
\hat c_k
\hat c_m
-
J
\sum_{jm}
(\widetilde I^2)_{jm}
\hat c_j^\dagger\hat c_m ,
\label{eq:normal-ordered-square}
\end{split}
\end{equation}
which is Eq.~\eqref{eq:hamiltonian_effective}. In the numerical calculations, the subdominant quadratic term is dropped.

\subsubsection*{2. Cavity dissipation}
\label{cavity-dissipation-single-mode}

Photon leakage through the cavity mirrors at rate $\kappa$ is described microscopically by the Lindblad jump operator $\hat L_{\rm c}=\sqrt{\kappa}\,\hat a$.
The same elimination of the cavity mode used above can be applied directly to $\hat L_{\rm c}$. 
Including the cavity linewidth, the equation of motion for $\hat a$ reads
\begin{equation}
\frac{\partial\hat a}{\partial t}
=
-\left(\rmi\Delta+\frac{\kappa}{2}\right)\hat a
-\rmi\Lambda\hat F\qquad\Longrightarrow\qquad\hat a
\simeq
-\frac{\Lambda\hat F}{\Delta-\rmi\kappa/2},
\end{equation}
since in the dispersive regime the cavity follows the fermionic dynamics adiabatically.
An overall phase of a Lindblad jump operator is immaterial. We can therefore write the effective cavity-loss channel as
\begin{equation}
\hat L_{\rm eff}
=
\sqrt{\kappa_{\rm eff}}\,\hat F,
\qquad\qquad
\kappa_{\rm eff}
=
\kappa\frac{\Lambda^2}{\Delta^2+\kappa^2/4}
\simeq
\kappa\left(\frac{\Lambda}{\Delta}\right)^2,
\label{eq:supp_cavity_jump}
\end{equation}
where the last expression holds for $|\Delta|\gg\kappa$.
Importantly, the overlaps $\widetilde I_{jk}$ entering $\hat F$ are precisely the same disordered couplings as in the Hamiltonian.
For example, the recycling term of the corresponding Lindblad dissipator is
\begin{equation}
\hat L_{\rm eff}\hat\rho\hat L_{\rm eff}^{\dagger}
=
\kappa_{\rm eff}
\sum_{jk\ell m}
\widetilde{I}_{jk}\widetilde{I}_{\ell m}\,
\hat c_j^\dagger\hat c_k\,
\hat\rho\,
\hat c_\ell^\dagger\hat c_m.
\label{eq:supp_cavity_dissipator}
\end{equation}
The coefficient tensor therefore factorizes as $\widetilde I_{jk}\widetilde I_{\ell m}$: cavity loss generates a single, collective nonlocal dephasing channel. This low-rank structure stands in contrast to the spontaneous-emission dissipator derived below.

\subsubsection*{3. Atomic spontaneous emission}

Atomic spontaneous emission corresponds, after elimination of the atomic excited states, to Rayleigh scattering of a drive photon into free space, and subsequent atomic recoil at momentum $k_0=2\pi/\lambda_\rmc$. 
A microscopic treatment couples
$\hat{\Psi}_{\rm g}^{\dagger}\hat{\Psi}_{\rm e}$ to the continuum of vacuum electromagnetic modes and traces out this continuum~\cite{pichler_nonequilibrium_2010, daley_quantum_2014}.
In this work, we assume isotropic scalar emission and, while tracing out the free-space continuum, we retain only the dissipative part of the field correlator and neglect its principal-value counterpart, the light-induced dipole–dipole interaction~\cite{lehmberg_radiation_1970}. 
Like the dissipator, this term is proportional to $\Gamma$ and vanishes in the closed-system limit. 
Within the model considered here, chaos is therefore induced by the jump operators alone.
The jump operator associated with emission along the unit vector $\bfn=
(\sin\theta\cos\varphi,\,
 \sin\theta\sin\varphi,\,
 \cos\theta)$
reads
\begin{equation}
    \hat L_{\rm a}(\bfn)
    =
    \sqrt{\Gamma}
    \int \rmd^2u\,
    \rme^{-\rmi\eta\bfn_\perp\cdot\bfu}
    \hat{\Psi}_{\rm g}^{\dagger}(\bfu)
    \hat{\Psi}_{\rm e}(\bfu) 
    =
    \sqrt{\Gamma_{\rm eff}}
    \sum_{jk}
    A_{jk}(\bfn,\eta)
    \hat c_j^\dagger\hat c_k,
\label{eq:supp_spontaneous_direction}
\end{equation}
where
$\bfn_\perp=(n_x,n_y)$
is the projection of the photon direction onto the atomic plane. 
Although the atomic motion is restricted to the $xy$ plane, photons are emitted over the full solid angle; recoil along the tightly confined transverse direction is frozen out.
In addition, $\eta=k_0x_0$ is the Lamb-Dicke parameter, $\Gamma_{\rm eff}=\Gamma\left(\Omega_\rmd/\Delta_{\rm da}\right)^2 = 4\Gamma\Lambda^2/\Omega^2$ is an effective spontaneous-emission rate, while 
\begin{equation}
    A_{jk}(\bfn,\eta)
    =
    \int \rmd^2u\,
    \rme^{-\rmi\eta\bfn_\perp\cdot\bfu}
    w_{\rm a}(\bfu)
    \phi_j^*(\bfu)
    \phi_k(\bfu),
    \label{eq:recoil_resolved_amplitude}
\end{equation}
is the recoil-resolved orbital amplitude with 
\begin{equation}
    w_{\rm a}(\bfu)
    =
    \frac{
    g_\rmd(\bfu)}
    {\Delta_{\rm da}(\bfu)/\Delta_{\rm da}}\simeq w(\bfu).
\end{equation}
To arrive at the right-hand side of Eq.~\eqref{eq:supp_spontaneous_direction} we performed adiabatic elimination of the atomic excited states, neglected the term proportional to $\Omega\hat{a}$ since the cavity is in the vacuum and $\Omega_\rmd\gg\Omega$~\cite{FerrariEtAl2026}, and expanded the fermionic field operator in single-particle wave functions.
Performing the sum over all possible angular directions and normalizing gives the pair-index Kossakowski matrix
\begin{equation}
\begin{split}
    K_{jk\ell m}
    =
    \int\frac{\rmd\Omega_{\bfn}}{4\pi}\,
    A_{jk}(\bfn,\eta)
    A_{\ell m}^{*}(\bfn,\eta)
    =
    \int \rmd^2u\,\rmd^2u'\,
    j_0\!\left(
    \eta|\bfu-\bfu'|
    \right)
    w_{\rm a}(\boldsymbol u)
    w_{\rm a}^{*}(\boldsymbol u')
    \phi_j^*(\bfu)
    \phi_k(\bfu)
    \phi_\ell(\bfu')
    \phi_m^*(\bfu'),
\label{eq:recoil-kossakowski}
\end{split}
\end{equation}
where the last equality follows from
\begin{equation}
    \int\frac{\rmd\Omega_{\bfn}}{4\pi}\rme^{-\rmi\eta\bfn_\perp\cdot(\bfu-\bfu')} 
    =
    \int_0^\pi
    \frac{\sin\theta\,\rmd\theta}{2}
    \mathcal J_0\!\left(
    \eta\sin\theta
    |\bfu-\bfu'|
    \right)
    =
    j_0\!\left(
    \eta|\bfu-\bfu'|
    \right),
\end{equation}
where $\JJ_0(x)$ is the Bessel function of first kind and $j_0(x) = \sin(x)/x$ the spherical Bessel function of first kind.
Equation~\eqref{eq:recoil-kossakowski} contains the recoil, heating, and inter-orbital dephasing generated by photon scattering. The same speckle-dependent detuning enters both
$K_{jk\ell m}$ and the Hamiltonian overlaps, so the dissipative and coherent couplings are correlated.
Defining
$\hat B_{jk}=\hat c_j^\dagger\hat c_k$,
the dissipator takes the compact form
\begin{equation}
\begin{split}
\mathcal D_{\rm a}\hat\rho
=
\Gamma_{\rm eff}
\sum_{jk\ell m}
K_{jk\ell m}
\bigg[\hat B_{jk}\hat\rho\hat B_{\ell m}^{\dagger}-\frac{1}{2}
\left\{
\hat B_{\ell m}^{\dagger}\hat B_{jk},
\hat\rho
\right\}
\bigg].
\label{eq:supp_spontaneous_tensor_dissipator}
\end{split}
\end{equation}
The Kossakowski matrix
$K_{jk\ell m}$
is Hermitian and positive semidefinite because it is the Gram matrix of the directional amplitudes in Eq.~\eqref{eq:recoil_resolved_amplitude}. It can therefore be diagonalized as
\begin{equation}
K_{jk\ell m}
=
\sum_{\alpha=1}^{R}
\lambda_\alpha
V_{jk}^{(\alpha)}
V_{\ell m}^{(\alpha)*},
\qquad\textrm{where}\qquad
R=\operatorname{rank}K .
\label{eq:recoil-factorization}
\end{equation}
Absorbing the eigenvalue into the coupling matrix, $M_{jk}^{(\alpha)}=\sqrt{\lambda_\alpha}\,V_{jk}^{(\alpha)}$
yields the jump operators used in the main text,
\begin{equation}
\hat L_{\rm a}^{(\alpha)}
=
\sqrt{\Gamma_{\rm eff}}
\sum_{jk}
M_{jk}^{(\alpha)}
\hat c_j^\dagger\hat c_k.
\label{eq:supp_spontaneous_final_jumps}
\end{equation}
We can take two limits to clarify the physical meaning of this decomposition. First, in the deep Lamb-Dicke regime $\eta\ll1$, $j_0\!\left(\eta|\bfu-\bfu'|\right)\simeq1$,
and the Kossakowski matrix factorizes as
\begin{equation}
K_{jk\ell m}
=
A_{jk}(\bfn,0)
A_{\ell m}^{*}(\bfn,0).
\end{equation}
It therefore has rank one. For the broad-mode profiles employed here, $A_{jk}(\bfn,0) = \widetilde I_{jk} = g_{jk}/\Lambda$,
and the spontaneous-emission dissipator reduces to the single collective jump
\begin{equation}
\hat L_{\rm a}^{(1)}
=
\sqrt{\Gamma_{\rm eff}}
\sum_{jk}
\widetilde I_{jk}
\hat c_j^\dagger\hat c_k.
\label{eq:supp_lamb_dicke_jump}
\end{equation}
For a spatially uniform drive and in the absence of speckle,
$\widetilde I_{jk}=\delta_{jk}$, so the jump operator is proportional to the total fermion number in a fixed-$Q$ sector. In the presence of disorder it remains a single, generally nonlocal dephasing channel.
Second, the leading recoil corrections can be obtained for $\eta\lesssim1$ by expanding Eq.~\eqref{eq:recoil_resolved_amplitude}:
\begin{equation}
\begin{split}
A_{jk}(\bfn,\eta)
\simeq
A_{jk}(\bfn,0)-\rmi\eta
\int \rmd^2u\,
(\bfn_\perp\cdot\bfu)
w_{\rm a}(\bfu)
\phi_j^*(\bfu)
\phi_k(\bfu).
\label{eq:supp_recoil_expansion}
\end{split}
\end{equation}
The second term is of order $\eta$ and, for harmonic-oscillator states, couples orbitals differing by one motional quantum. It therefore generates incoherent orbital-changing transitions, diffusion, and heating. For scalar three-dimensional emission, the associated rate scales as $\Gamma_{\rm heat}\sim\frac{2}{3}\Gamma_{\rm eff}\eta^2$ where $\langle k_\perp^2\rangle/k_0^2=2/3$.
Thus, increasing $\eta$ resolves the atomic motion during a scattering event, destroys the rank-one structure of the deep Lamb-Dicke limit, and activates a collection of nonlocal dephasing channels
$\hat L_{\rm a}^{(\alpha)}$
used in Eq.~\eqref{eq:lindblad}.

\subsubsection*{4. Experimental challenges}
\label{sec:realizability}

Our protocol for dissipative SYK physics solves two severe constraints of quantum simulation of chaotic models with cavity QED devices: no complicated time-dependent protocol as in Refs.~\cite{Baumgartner2024, Baumgartner2025, FerrariEtAl2026} is required, nor dissipation challenges the many-body coherence, as detailed in Ref.~\cite{FerrariEtAl2026}, since dissipation is what we harness to create chaotic behavior.
It \textit{does not}, however, avoid two remaining experimental challenges, arising in any driven-dissipative cavity QED platform: the fermionic trap Hamiltonian and dissipation-induced leakage to higher trap levels, a process commonly referred to as heating.
For simplicity, throughout this work we have described the fermionic orbitals by a truncated basis of $N$ harmonic-oscillator eigenstates. This choice does not affect the spectral properties of the Liouvillian in Eq.~\eqref{eq:lindblad}, which is a well-defined generator on the retained $N$-orbital space. 
A direct implementation in a harmonic trap, however, faces the two aforementioned issues.
First, we have neglected the single-particle trap Hamiltonian $\hat H_{\rm t}=\int \rmd^2r\,\hat\Psi_g^\dagger(\mathbf r)\hat h_{\rm t}\hat\Psi_g(\mathbf r)$ of Eq.~\eqref{eq:supp_rotating_hamiltonian}, which is diagonal in the orbitals with level spacing $\omega_{\rm t}$. 
This term is trivially integrable and at small oscillator lengths, $\omega_{\rm t}\gg J$, thus hindering the development of quantum chaos.
Second, spontaneous emission is not confined to the retained orbitals. A recoil kick takes an atom in orbital $j$ out of the manifold with probability 
\begin{equation}
    p^{\rm out}_j=1-\sum_{k\le N}\big|\langle k|\rme^{-\rmi k_0\mathbf n\cdot\mathbf r}|j\rangle\big|^2,
\end{equation}
and deposits on average the energy $\tfrac23E_R=\tfrac{\eta^2}{3}\omega_{\rm t}$ in the in-plane motion. 
This leads to heating and, eventually, atom loss. 
Since the scattered photon carries the energy away, an energy gap alone does not prevent these transitions.
The two issues share a single origin: the retained orbitals do not form an isolated, flat band. 
In a harmonic trap, the same scale $\omega_{\rm t}$ sets both the in-band dispersion, which should be small compared with $J$, and the leakage, which grows with $\eta^2=2E_{\rm R}/\omega_{\rm t}$. 
 
Both issues can be mitigated at once by replacing the harmonic orbitals with the lowest Landau level (LLL) of a two-dimensional gas in a synthetic magnetic field~\cite{dalibard_colloquium_2011, cooper_optical_2011}. 
The field has cyclotron frequency $\omega_{\rm B}$ and magnetic length $\ell=\sqrt{1/m_{\rm at}\omega_{\rm B}}$. 
Writing $\mathbf r=\mathbf R+\boldsymbol\rho_\rmc$, where $\mathbf R$ is the guiding centre with $[R_x,R_y]=-i\ell^2$, the single-particle spectrum in a weak harmonic confinement $\omega_{\rm t}\ll\omega_{\rm B}$ is
\begin{equation}
    E_{n,m}\simeq\omega_{\rm B}\Big(n+\tfrac12\Big)+\frac{\omega_{\rm t}^2}{\omega_{\rm B}}\,m ,
\end{equation}
with Landau-level index $n$ and guiding-centre index $m$.
The microscopic derivation of the cavity QED Hamiltonian and jump operators carries over unchanged, with three replacements:
\begin{itemize}
    \item the harmonic orbitals $\phi_j$ of Eq.~\eqref{eq:supp_ho_wavefunction} become the $N$ LLL orbitals $\phi_j(\mathbf r)\propto(z/\ell)^{j}e^{-|z|^2/4\ell^2}$ ($n=0$, $m=j=0,\dots,N-1$, $z=x+\rmi y$), with the number of retained orbitals $N$ now fixed by the cloud area ($N\simeq A/2\pi\ell^2$) rather than by a cutoff;
    \item the length unit becomes $x_0\to\ell$;
    \item the Lamb--Dicke parameter becomes $\eta=k_0x_0\to\eta_{\rm LL}=k_0\ell$.
\end{itemize}
In particular,
\begin{equation}
\hat F=\int d^2r\,w(\mathbf r)\,\hat\Psi_g^\dagger(\mathbf r)\hat\Psi_g(\mathbf r)\qquad\Longrightarrow\qquad P_0\hat FP_0=\sum_{jk}\tilde I_{jk}\,\hat c_j^\dagger\hat c_k,
\end{equation}
with $w$ and $\tilde I_{jk}$ evaluated on LLL orbitals. 
The in-band jump operators are $P_0\hat L_a(\mathbf n)P_0=\sqrt{\Gamma_{\rm eff}}\sum_{jk}A_{jk}(\mathbf n,\eta_{\rm LL})\,\hat c_j^\dagger\hat c_k$, with $A_{jk}$ as in Eq.~\eqref{eq:recoil_resolved_amplitude}.
The speckle that generates $\Delta_{\rm da}(\mathbf r)$ has correlation length $\xi$, and hence wavevector cutoff $k_w\sim1/\xi$. 
Projected onto Landau levels, plane waves obey the Girvin--MacDonald--Platzman (GMP) relations~\cite{girvin_magneto-roton_1986}
\begin{equation}
    P_n \rme^{\rmi\mathbf q\cdot\mathbf r}P_0=G_{n0}(\mathbf q\ell)\,\rme^{\rmi\mathbf q\cdot\mathbf R},\qquad\qquad
    |G_{n0}|^2=\frac{1}{n!}\Big(\frac{q^2\ell^2}{2}\Big)^{n}\rme^{-q^2\ell^2/2}.
\label{eq:GMP}
\end{equation}
A single scale, the cyclotron gap $\omega_{\rm B}$, then controls everything that spoils the projected model:
 
\textit{(i) Trap Hamiltonian.} Within the LLL ($n=0$), $\hat H_{\rm t}$ reduces to a dispersion $\omega_{\rm t}^2/\omega_{\rm B}$ per guiding-centre state. It is suppressed by $\omega_{\rm t}/\omega_{\rm B}$ with respect to the harmonic case.
 
\textit{(ii) Leakage.} By Eq.~\eqref{eq:GMP}, a recoil kick factorizes into a unitary in-band part $\rme^{\rmi\mathbf q\cdot\mathbf R}$ and an amplitude $\rme^{-q^2\ell^2/4}$. Averaging over emission directions ($q=k_0\sin\theta$), the out-of-band probability per scattering event is
\begin{equation}
\varepsilon=1-\big\langle \rme^{-q^2\ell^2/2}\big\rangle_{\mathbf n}\simeq\frac{(k_0\ell)^2}{3}=\frac{2E_R}{3\omega_{\rm B}}.
\label{eq:eps}
\end{equation}
Here, in contrast to a generic trap, the gap \emph{does} suppress leakage. Increasing $\omega_{\rm B}$ shrinks $\ell$ below the optical wavelength, so the recoil can no longer resolve, and hence excite, the cyclotron motion.
 
\textit{(iii) Virtual excitations.} Eliminating the cavity with the atoms in the LLL gives Landau-level-resolved denominators,
\begin{equation}
\hat H_{\rm cav}=-\sum_{n\ge0}J_n\,P_0\hat FP_n\hat FP_0,\qquad J_n=\frac{\Lambda^2}{\Delta+n\omega_{\rm B}},
\label{eq:Hcav}
\end{equation}
with $\Lambda$ as in Eq.~\eqref{eq:supp_single_mode_overlap}, $\Delta=\omega_c-\omega_d$ the cavity detuning, and $J_0=J=\Lambda^2/\Delta$ as in Eq.~(S10).
The $n=0$ term, $-J(P_0\hat FP_0)^2$, is the rank-one Hamiltonian of the main text. For $n\ge1$, the second $\hat F$ must return the same atom to the LLL. 
These terms are therefore one-body, and suppressed by $\Delta/\omega_{\rm B}$. 
Any uncompensated fraction, of amplitude $U_0$, is harmless at leading order. 
For broad modes ($g_d\simeq g_c\simeq1$) it is proportional to $w(\mathbf r)$, so its in-band part is $\propto P_0\hat FP_0$ and commutes with $\hat H_{\rm cav}$. 
Only its virtual Landau-level mixing, $\propto U_0^2/\omega_{\rm B}$, survives.
 
Altogether, every correction to the integrable form $-J(P_0\hat FP_0)^2$ (up to a commuting term $\propto P_0\hat FP_0$) is a one-body potential for the guiding centres. Its bandwidth is
\begin{equation}
\delta_{\rm res}\simeq\frac{\Delta}{\omega_{\rm B}}\frac{(k_w\ell)^2}{2}\,J+\frac{U_0^2(k_w\ell)^2}{2\omega_{\rm B}}+N\,\frac{\omega_{\rm t}^2}{\omega_{\rm B}},
\label{eq:dres}
\end{equation}
where $N$ is the number of occupied LLL orbitals. Equations~\eqref{eq:eps} and~\eqref{eq:dres} show that the trap dispersion, the leakage and the virtual corrections all decrease as $1/\omega_{\rm B}$. The same large gap flattens the band, closes the leakage channel and suppresses inter-level processes.
Within the LLL, projected density operators do not commute:
\begin{equation}
\big[\rme^{\rmi\mathbf k\cdot\mathbf R},\rme^{\rmi\mathbf k'\cdot\mathbf R}\big]=2\rmi\sin\!\Big(\frac{\ell^2}{2}\,\mathbf k\times\mathbf k'\Big)\,\rme^{\rmi(\mathbf k+\mathbf k')\cdot\mathbf R}.
\end{equation}
The projected speckle operator $P_0\hat FP_0$ and the projected recoil operators $P_0\hat L_a(\mathbf n)P_0$ therefore fail to commute, as in the truncated model. This is a property of the physical band rather than of a numerical cutoff. The parameter $\eta_{\rm LL}\equiv k_0\ell=\sqrt{2E_R/\omega_{\rm B}}$ plays the role of $\eta=k_0x_0=\sqrt{2E_R/\omega_{\rm t}}$. 
For $\eta_{\rm LL}\to0$, every jump operator reduces to a single collective channel $\propto P_0\hat FP_0$, recovering the integrable corner of the End Matter. 
The non-commutativity $[P_0\hat FP_0,P_0\hat L_aP_0]$ is of order $(k_w\ell)\,\eta_{\rm LL}/2$, i.e.\ \emph{first} order in $\eta_{\rm LL}$, whereas leakage is \emph{second} order [c.f. Eq.~\eqref{eq:eps}]. 
In conclusion, such a protocol might suppress the two remaining experimental issues: the trivially integrable trapping frequency and recoil-induced heating. 
Its main requirement of this protocol is a large cyclotron gap, $\omega_{\rm B}$ of several $E_R$ ($\eta_{\rm LL}\lesssim1$). 
A more quantitative analysis, including heating from the gauge-field lasers, is beyond the scope of this work.

\subsection*{B. Multimode cavity}

\subsubsection*{1. Hamiltonian}

The derivation for a multimode cavity closely follows the single-mode case discussed in the previous section. The only essential modification is the presence of several transverse cavity modes, which independently mediate coherent interactions and photon-loss channels~\cite{Uhrich2023,Baumgartner2024}. Throughout this subsection we neglect atomic spontaneous emission in order to isolate the effect of cavity losses.
We retain $M$ transverse cavity modes, labeled by
$\mu=(n_x,n_y)$, and define their transverse order as $m_\mu=n_x+n_y$.
For an approximately equally spaced transverse-mode spectrum, the cavity-drive detunings are
\begin{equation}
    \Delta_\mu=\Delta+m_\mu\delta\omega,
    \qquad
    \delta\tilde{\omega}=\frac{\delta\omega}{\Delta},
    \qquad
    q_\mu=\frac{\Delta}{\Delta_\mu}=\frac{1}{1+m_\mu\delta\tilde{\omega}}.
\label{eq:supp_multimode_detunings}
\end{equation}
More generally, $\mu$ may be understood as a composite transverse-mode index, with $\Delta_\mu$ fixed by the corresponding cavity spectrum.
After eliminating the atomic excited states as in the previous section, the disorder-dependent overlap associated with mode $\mu$ is
\begin{equation}
    \widetilde I_{jk}^{(\mu)}
    =
    \int \rmd^2u\,
    \frac{
    g_\rmd(\bfu)
    g_\mu^*(\bfu)
    \phi_j^*(\bfu)
    \phi_k(\bfu)}
    {\Delta_{\rm da}(\bfu)/\Delta_{\rm da}},
\label{eq:supp_multimode_overlap}
\end{equation}
where $g_\mu(\bfu)$ is the transverse profile of mode $\mu$. In contrast to the single-mode calculation, we retain the transverse dependence of $g_\mu(\bfu)$, while continuing to take the drive broad, $g_\rmd(\bfu)\simeq1$. This dependence is essential: if all $g_\mu$ were spatially uniform or proportional over the atomic cloud, the matrices $g_{jk}^{(\mu)}$ would be proportional, reducing both the multimode interaction and the cavity-loss dissipator to the single-mode, low-rank structure~\cite{Uhrich2023}. 
Introducing
\begin{equation}
\Lambda_\mu
=
\frac{\Omega_\mu\Omega_\rmd}{2\Delta_{\rm da}},
\qquad
g_{jk}^{(\mu)}
=
\Lambda_\mu\widetilde I_{jk}^{(\mu)},
\label{eq:supp_multimode_gjk}
\end{equation}
the photon-assisted Hamiltonian becomes
\begin{equation}
\hat H
=
\sum_{\mu=1}^{M}
\left[
\Delta_\mu\hat a_\mu^\dagger\hat a_\mu
+
(\hat a_\mu^\dagger+\hat a_\mu)\hat F_\mu
\right],
\qquad
\hat F_\mu
=
\sum_{jk}
g_{jk}^{(\mu)}
\hat c_j^\dagger\hat c_k,
\label{eq:supp_multimode_hamiltonian}
\end{equation}
which is Eq.~\eqref{eq:hamiltonian_multimode} of the main text. The same speckle realization enters all
$\widetilde I_{jk}^{(\mu)}$
through the disordered atomic detuning, while the distinct transverse profiles
$g_\mu(\bfu)$
generate different coupling matrices. For the range of modes considered here, mode-to-mode variations of $\Omega_\mu$ can be absorbed into
$g_{jk}^{(\mu)}$; when these variations are negligible, one may set
$\Lambda_\mu=\Lambda$.
We next eliminate the cavity modes in the dispersive regime (i.e., $|\Delta_\mu|\gg\|\hat F_\mu\|$
for all relevant $\mu$) by means of a Schrieffer-Wolff transformation~\cite{Uhrich2023}, obtaining
\begin{equation}
\hat H_{\rm eff}
=
-\sum_{\mu=1}^{M}
\frac{\hat F_\mu^2}{\Delta_\mu}
=
-\sum_{\mu=1}^{M}
\frac{1}{\Delta_\mu}
\Big[
\sum_{jk}
g_{jk}^{(\mu)}
\hat c_j^\dagger\hat c_k
\Big]^2.
\label{eq:supp_multimode_effective_compact}
\end{equation}
Different cavity modes therefore contribute additively to the effective interaction; cross terms between distinct photonic modes vanish upon projection onto the vacuum sector.
Normal ordering Eq.~\eqref{eq:supp_multimode_effective_compact} separates a one-body contribution from the desired two-body interaction:
\begin{equation}
\hat H_{\rm eff}
=
-\sum_{\mu,jkm}
\frac{
g_{jk}^{(\mu)}
g_{km}^{(\mu)}}
{\Delta_\mu}
\hat c_j^\dagger\hat c_m+
\sum_{\mu,jk\ell m}
\frac{
g_{jk}^{(\mu)}
g_{\ell m}^{(\mu)}}
{\Delta_\mu}
\hat c_j^\dagger
\hat c_\ell^\dagger
\hat c_k
\hat c_m.
\label{eq:supp_multimode_normal_order}
\end{equation}
The first term is the multimode analogue of the one-body term discussed in Sec.~\ref{sec:single-mode-cavity} and can be compensated or made subleading relative to the quartic interaction~\cite{Uhrich2023,Baumgartner2024}. Because of fermionic antisymmetry, only the antisymmetrized component of the four-index coupling contributes. The interacting part can thus be written as
\begin{equation}
\hat H_{\rm eff}^{(4)}
=
\sum_{jk\ell m}
g_{j\ell km}\,
\hat c_j^\dagger
\hat c_\ell^\dagger
\hat c_k
\hat c_m,
\qquad 
g_{j\ell km}
=
\frac{1}{2}
\sum_{\mu=1}^{M}
\frac{
g_{jk}^{(\mu)}g_{\ell m}^{(\mu)}
-
g_{jm}^{(\mu)}g_{\ell k}^{(\mu)}}
{\Delta_\mu}.
\label{eq:supp_multimode_H4}
\end{equation}
For approximately mode-independent Raman amplitudes,
$\Lambda_\mu=\Lambda$, the same coupling can be expressed using the overlap and mode-weight notation as
\begin{equation}
g_{j\ell km}
=
\frac{J}{2}
\sum_{\mu=1}^{M}
q_\mu
\left[
\widetilde I_{jk}^{(\mu)}
\widetilde I_{\ell m}^{(\mu)}
-
\widetilde I_{jm}^{(\mu)}
\widetilde I_{\ell k}^{(\mu)}
\right].
\label{eq:supp_multimode_four_index_overlap}
\end{equation}
The right-hand side of Eq.~\eqref{eq:supp_multimode_H4} makes the low-rank structure of the cavity construction explicit: each photonic mode contributes one separable product of two coupling matrices, while summing over many modes progressively increases the rank of the effective four-fermion interaction. This is the mechanism by which the multimode cavity approaches the dense random interaction structure of the SYK model~\cite{Uhrich2023}.
The relative weight of mode $\mu$ in Eq.~\eqref{eq:supp_multimode_effective_compact} is controlled by $q_\mu$.
Thus, $\delta\tilde{\omega}\gg1$ suppresses all but the fundamental cavity mode and recovers the rank-one interaction of Sec.~\ref{sec:single-mode-cavity}. Conversely, decreasing $\delta\tilde{\omega}$ brings an increasing number of transverse modes into the dynamics, thereby increasing the effective interaction rank. In practice, $M$ denotes the set of modes retained in the numerical calculation, whereas $\delta\tilde{\omega}$ controls their effective participation through the weights $q_\mu$.

\subsubsection*{2. Cavity dissipation}
Each cavity mode loses photons through the mirrors and is associated with the microscopic jump operator $\hat L_{{\rm c},\mu}=\sqrt{\kappa_\mu}\,\hat a_\mu$.
The adiabatic elimination proceeds independently for each photonic mode. 
Including each cavity linewidth, the equation of motion for $\hat a_\mu$ reads
\begin{equation}
\frac{\partial{\hat a}_\mu}{\partial t} = -\left(\rmi\Delta_\mu+\frac{\kappa_\mu}{2}\right)\hat a_\mu-\rmi\hat F_\mu
\qquad\Longrightarrow\qquad
\hat a_\mu
\simeq
-\frac{\hat F_\mu}
{\Delta_\mu-\rmi\kappa_\mu/2},
\end{equation}
since, in the dispersive regime, each cavity field follows the fermionic dynamics adiabatically.
The $\mu$-th single-photon jump operator therefore becomes, after removing the common phase associated with the complex cavity susceptibility,
\begin{equation}
\hat L_{\rm eff}^{(\mu)}
=
\sqrt{\kappa_{\rm eff}^{(\mu)}}
\sum_{jk}
\widetilde{I}_{jk}^{(\mu)}
\hat c_j^\dagger\hat c_k,
\qquad
\kappa_{\rm eff}^{(\mu)}
=
\kappa_\mu\frac{\Lambda_\mu^2}
{\Delta_\mu^2+\kappa_\mu^2/4}.
\label{eq:supp_multimode_effective_jump}
\end{equation}
For $|\Delta_\mu|\gg\kappa_\mu$,
$\kappa_{\rm eff}^{(\mu)}
\simeq\kappa_\mu(\Lambda/\Delta_\mu)^2$.
If the relevant transverse modes are nearly degenerate and have approximately equal linewidths,
$\Delta_\mu\simeq\Delta$ and
$\kappa_\mu\simeq\kappa$, one may further use the common coefficient $\kappa_{\rm eff}=\kappa\frac{\Lambda^2}{\Delta^2+\kappa^2/4}$,
which yields the notation employed in the main text.
Keeping the finite cavity linewidth also replaces the coefficient
$1/\Delta_\mu$
in Eq.~\eqref{eq:supp_multimode_effective_compact} by
\begin{equation}
\frac{1}{\Delta_\mu}
\longrightarrow
\frac{\Delta_\mu}
{\Delta_\mu^2+\kappa_\mu^2/4}.
\end{equation}
This correction is negligible in the dispersive regime assumed throughout the manuscript.
The resulting fermionic master equation, in the absence of spontaneous emission, is therefore
\begin{equation}
\frac{\partial\hat\rho}{\partial t}
=
-\rmi[\hat H_{\rm eff},\hat\rho]
+
\sum_{\mu=1}^{M}
\mathcal D[\hat L_{\rm eff}^{(\mu)}]\hat\rho.
\label{eq:supp_multimode_lindblad}
\end{equation}
For real $\widetilde I_{jk}^{(\mu)}$, each
$\hat L_{\rm eff}^{(\mu)}$
is Hermitian after removal of the unphysical common phase and therefore describes collective fermionic dephasing. Crucially, the multimode cavity produces one such nonlocal channel for each participating photonic mode. The same parameter
$\delta\tilde{\omega}$
that increases the rank of the coherent interaction consequently also increases the rank of the cavity-loss dissipator: for large
$\delta\tilde{\omega}$
the system reduces to the single collective channel of the previous section, whereas for small
$\delta\tilde{\omega}$
a collection of distinct dephasing operators participates in the dynamics. Their couplings are not independent of the Hamiltonian disorder; each
$\hat L_{\rm eff}^{(\mu)}$
contains precisely the matrix
$\widetilde I_{jk}^{(\mu)}$
entering the corresponding contribution to Eq.~\eqref{eq:supp_multimode_effective_compact}.

\subsection*{C. Physical parameters}
The physical parameters used in all numerical simulations of this paper, for both the single- and multimode setup, are reported in Table~\ref{tab:physical-parameters}.
The details of the speckle construction and associated parameters can be found in Ref.~\cite{FerrariEtAl2026}.

\begin{table*}[t]
    \centering
    \caption{Parameters used for the single-mode spontaneous-emission and
    multimode cavity-loss models in
    Figs.~\ref{fig:atomic_spontaneous_emission},
    \ref{fig:single-mode-dynamics}, and
    \ref{fig:cavity_dissipation}.}
    \label{tab:physical-parameters}
    \setlength{\tabcolsep}{10pt}
    \renewcommand{\arraystretch}{1.08}
    \begin{tabular}{@{}lcc@{}}
        \toprule
        & \multicolumn{1}{c}{Single-mode}
        & \multicolumn{1}{c}{Multimode} \\
        \cmidrule(lr){2-2}
        \cmidrule(l){3-3}
        Parameter
        & Spontaneous emission
        & Cavity loss \\
        \midrule

        Main sector, $(N,Q)$
        & $(10,4)$
        & $(10,4)$ \\

        Inset sectors, $(N,Q)$
        & $(6,3),(8,4),(10,4)$
        & $(6,3),(8,4),(10,4)$ \\

        Inset control parameter
        & $\eta=2$
        & $\delta\tilde{\omega}=0.01$ \\

        \addlinespace[3pt]

        Number of cavity modes, $M$
        & $1$
        & $301$ \\

        Cavity--drive detuning, $\Delta/2\pi$
        & $\SI{1}{\mega\hertz}$
        & $\SI{1}{\mega\hertz}$ \\

        Cavity decay rate, $\kappa/2\pi$
        & $\SI{0.2}{\mega\hertz}$
        & $\SI{0.2}{\mega\hertz}$ \\

        \addlinespace[3pt]

        Spontaneous-emission ratio,
        $\Gamma_{\rm eff}/J$
        & $1$
        & -- \\

        Cavity-loss ratio, $\kappa/\Delta$
        & $0.2$
        & $0.2$ \\

        \addlinespace[3pt]

Cavity wavelength for ${}^6\rm Li$, $\lambda_{\rm c}$
& $\SI{671}{\nano\meter}$
& $\SI{671}{\nano\meter}$ \\

        Oscillator length, $x_0$
        & $\{10.7,42.7,106.8,213.6\}\,\si{\nano\meter}$
        & $\SI{5.8}{\micro\meter}$ \\

        \bottomrule
    \end{tabular}
\end{table*}

\section{Spectral symmetries of the Liouvillian}

\subsection*{A. Class \texorpdfstring{$\textrm{BDI}^\dagger$}{BDI†} and Hermitianization}

The symmetries of the Liouvillians fix the appropriate random-matrix references and hence the
quantitative predictions for the $\DSFF$, $\sSFF$, and CSR
~\cite{hamazaki_universality_2020,sa_symmetry_2023,
kawabata_symmetry_2023}.
Hermiticity preservation
implies a modular conjugation symmetry in each fixed-$Q$ block of the Liouvillian, while,
in the vectorized basis, the real-symmetric Hamiltonian and
jump matrices also imply transposition symmetry:
\begin{equation}
    \mathcal{J}\LL\mathcal{J}^{-1}=\LL,\qquad
    \mathcal{J}^{2}=1,\qquad
    \LL^\top=\LL ,
    \label{eq:liouvillian_symmetries}
\end{equation}
where $\mathcal{J}$ maps an operator to its adjoint.
The last relation is $\mathrm{TRS}^{\dagger}$ with square $+1$; together,
these constraints place the full block in class
$\mathrm{BDI}^{\dagger}$~\cite{sa_symmetry_2023,
kawabata_symmetry_2023}.
Away from the real axis, modular conjugation only pairs $\lambda_j$ with
the macroscopically separated $\lambda^{*}_j$, whereas transposition
symmetry constrains local correlations.
The generic complex bulk therefore has $\mathrm{AI}^{\dagger}$
universality~\cite{hamazaki_universality_2020,sa_symmetry_2023}.
Accordingly, the bulk CSR and the filtered $\DSFF$ at
$\phi=\pi/4$ are compared with Gaussian complex-symmetric matrices
processed identically to the spectra of the single and multimode models. The excluded real-axis subset
retains the full $\mathrm{BDI}^{\dagger}$ structure.
For singular values, which are not invariant under scalar shifts, we
instead center and Hermitianize the Liouvillian,
\begin{equation}
    \overline{\LL}
    =\LL-\frac{\Tr \LL}{\DD}\mathds{1},
    \qquad
    \mathcal{H}_{\overline{\LL}}
    =
    \begin{pmatrix}
        0 & \overline{\LL}\\
        \overline{\LL}^{\dagger} & 0
    \end{pmatrix},
    \qquad
    \operatorname{spec}\mathcal{H}_{\overline{\LL}}
    =\{\pm\sigma_j\}.
    \label{eq:liouvillian_hermitianization}
\end{equation}
The centering shift is real and preserves both symmetries.
The singular-value classification then maps
$\mathrm{BDI}^{\dagger}$ to Hermitian AI
~\cite{kawabata_singular-value_2023}.

\subsection*{B. Unfolded \texorpdfstring{$\sSFF$}{σSFF} as an instance of folded GOE universality class}

\begin{figure}[t] 
    \centering \includegraphics[width=0.9\textwidth]{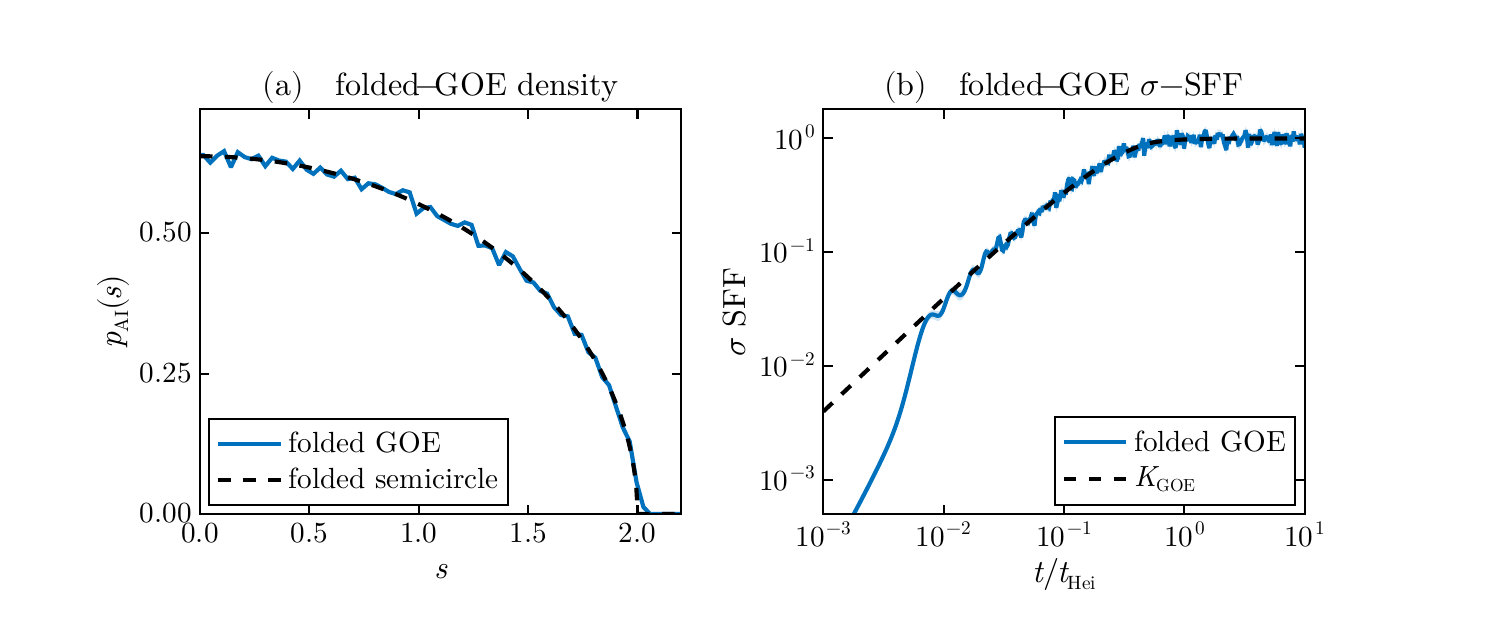} \caption{Numerical validation of the folded-GOE prediction in Eq.~\eqref{eq:folded_goe_sff}. (a) Singular-value density of $R=128$ trace-centered GOE matrices of size $N=400$, compared with the folded semicircle. (b) Unfolded connected $\sSFF$ (blue) compared with $K_{\rm GOE}(2|\tilde t|)$ (dashed).}
    \label{fig:f_goe}
\end{figure}

The AI reduction yields a signed GOE spectrum, whereas the $\sSFF$
retains only its absolute values.
The appropriate Gaussian reference is therefore
$\sigma_j=|E_j|$, with $\{E_j\}$ the spectrum of a trace-centered GOE
matrix. We refer to this ensemble as folded GOE (f-GOE).
At a singular value $\sigma>0$, folding combines the GOE eigenvalues lying locally around $+\sigma$ and $-\sigma$.
These two spectral regions are asymptotically independent and each carries half of
the folded density, so the time argument of the GOE form factor
is doubled.
Using the standard connected GOE form factor $K_{\rm GOE}$~\cite{Mehta2004}, for $\tilde t=t/t_{\rm Hei}$ and a connected plateau normalized to
unity,
\begin{equation}
\begin{split}
    K_{\mathrm{f\text{-}GOE}}(\tilde t)
    =K_{\mathrm{GOE}}(2|\tilde t|)
    =
    \begin{cases}
    4|\tilde t|-2|\tilde t|\ln\!\left(1+4|\tilde t|\right),
        & |\tilde t|<\tfrac{1}{2},\\[1mm]
    2-2|\tilde t|\ln\!\left(
        \dfrac{4|\tilde t|+1}{4|\tilde t|-1}
        \right),
        & |\tilde t|\geq\tfrac{1}{2}.
    \end{cases}
    \label{eq:folded_goe_sff}
\end{split}
\end{equation}
This factor of two fixes both the universal ramp and the Heisenberg
scale of the unfolded $\sSFF$
~\cite{haake_quantum_2018,kawabata_singular-value_2023}.
Equation~\eqref{eq:folded_goe_sff} is valid only in the bulk: near
$\sigma=0$ the two branches meet and hard-edge correlations become
relevant.
We therefore exclude the near-zero singular values and apply the same
bulk selection and unfolding to the single-mode, multimode and reference spectra.
We verify the folded-GOE prediction and the associated factor-of-two rescaling directly in Fig.~\ref{fig:f_goe}. The singular-value density follows the folded semicircle, while the connected unfolded $\sSFF$ agrees with $K_{\rm GOE}(2|\tilde t|)$ over the full ramp-to-plateau crossover, without fitted parameters. This agreement also confirms that the finite-size unfolding procedure applied to the reference ensemble reproduces the analytical f-GOE prediction.

\section{Spectral Form Factors}
\subsection*{A. Review of the general definitions}
\label{sec:sff-continuations}
For a Hermitian Hamiltonian $\hat H\ket{n}=E_n\ket{n}$, the
infinite-temperature SFF has two equivalent representations: as a spectral
sum over the Hamiltonian eigenvalues and as the return probability of the
infinite-temperature coherent Gibbs state
\(\ket{\Psi}=D^{-1/2}\sum_n\ket{n}\), obtained from
\(\ket{\Psi_\beta}\propto\sum_n \rme^{-\beta E_n/2}\ket{n}\) at \(\beta=0\):
\begin{equation}
 \SFF(t)=\frac{1}{D^2}\left|\Tr \rme^{-\rmi t\hat H}\right|^2
 =\frac{1}{D^2}\sum_{m,n}\rme^{-\rmi t(E_m-E_n)}
 =\left|\bra{\Psi}\rme^{-\rmi t\hat H}\ket{\Psi}\right|^2,
 \qquad
 \ket{\Psi}=\frac{1}{\sqrt D}\sum_n\ket{n}.
 \label{eq:sff-three-equivalent}
\end{equation}
This standard identity follows because the same Hermitian generator
determines both the spectrum and the unitary
evolution~\cite{DelCampo2017,xu_thermofield_2021,
matsoukas-roubeas_quantum_2024}.
For a Liouvillian with eigenvalues
$\lambda_\alpha=x_\alpha+\rmi y_\alpha$, the DSFF retains the spectral
construction by treating the spectrum as a two-dimensional point
process~\cite{li_spectral_2021},
\begin{equation}
 Z_\lambda(t,s)=\sum_{\alpha=1}^{\DD}
 e^{\rmi(tx_\alpha+sy_\alpha)},
 \qquad
 \DSFF(t,s)=\frac{1}{\DD^2}
 \left\langle\left|Z_\lambda(t,s)\right|^2\right\rangle ,
 \label{eq:eigenvalue-dsff}
\end{equation}
where angular brackets denote disorder averaging. Writing
$(t,s)=|\tau|(\cos\phi,\sin\phi)$, a fixed-$\phi$ cut is the ordinary
SFF of the complex spectrum projected along that direction. Throughout this section
we take $\phi=\pi/4$, as in the main text~\cite{li_spectral_2021}.
The return-probability continuation is instead the channel fidelity.
For a diagonalizable $\LL$,
\begin{equation}
 \SFF_{\rm fid}(t)
 =\vbra\hat\rho_\Psi|\rme^{\LL t}|\hat\rho_\Psi\vket
 =\sum_\alpha \rme^{\lambda_\alpha t}
 \vbra\hat\rho_\Psi|\hat\eta_\alpha\vket\vbra \hat\sigma_\alpha|\hat\rho_\Psi\vket,
 \qquad
 \hat\rho_\Psi=\ket{\Psi}\!\bra{\Psi},
 \label{eq:fidelity-sff}
\end{equation}
where $\vbra A|B\vket=\Tr(A^\dagger B)$ and
$\vbra \hat\sigma_\alpha|\hat\eta_\beta\vket=\delta_{\alpha\beta}$%
~\cite{xu_thermofield_2021,matsoukas-roubeas_quantum_2024}, and $|\hat{O}\vket$ denotes the vectorization of the operator $\hat{O}$.
Equations~\eqref{eq:eigenvalue-dsff} and \eqref{eq:fidelity-sff} are
therefore not equivalent for open systems: the DSFF places both $x_\alpha$
and $y_\alpha$ in phases and probes eigenvalue-pair correlations,
whereas the physical propagator contains the damping factor
$\rme^{x_\alpha t}$ and eigenmode-dependent weights.
Motivated by this inequivalence, we turn to singular-value diagnostics, which have recently been used to characterize quantum chaos in non-Hermitian Hamiltonians~\cite{roccati_diagnosing_2024,nandy_probing_2025,prasad_assessment_2025}.
The associated singular form factor is obtained from the Hermitian operator constructed from the singular values, as we now detail. 
All terms in our model conserve fermion number, so we restrict $\LL$ to the symmetry block with $Q$ particles. 
The resulting Hilbert space has dimension $D=\binom{N}{Q}$, and the corresponding operator space has dimension $\DD=D^2$. In what follows, $\LL$ denotes this $\DD\times\DD$ block. 
Since singular values are not invariant under scalar shifts, we follow Ref.~\cite{kawabata_singular-value_2023} and remove its scalar component before taking the SVD,
\begin{equation}
 \overline{\LL}=\LL-c\mathds{1}_{\DD},
 \qquad
 c=\frac{\operatorname{Tr}_{\rm HS}\LL}{\DD},
 \qquad
 \operatorname{Tr}_{\rm HS}\overline{\LL}=0.
 \label{eq:fixed-filling-centered-liouvillian}
\end{equation}

Writing
\begin{equation}
 \overline{\LL}=\sum_{j=1}^{\DD}\sigma_j|u_j\vket\vbra v_j|,
 \qquad
 S\equiv\sqrt{\overline{\LL}^{\dagger}\overline{\LL}}
 =\sum_{j=1}^{\DD}\sigma_j|v_j\vket\vbra v_j|,
 \label{eq:sigma-operator}
\end{equation}
reduces the construction to the positive Hermitian operator $S$, whose
eigenvalues are the singular values $\sigma_j\geq0$. Its ordinary
infinite-temperature SFF is
\begin{equation}
 \begin{split}
 Z_\sigma(t)
 =\operatorname{Tr}\rme^{-\rmi t\sqrt{\overline{\LL}^{\dagger}\overline{\LL}}}
 =\sum_{j=1}^{\DD}\rme^{-\rmi t\sigma_j},\qquad
 \sSFF(t)
 =\frac{1}{\DD^2}
 \left\langle\left|Z_\sigma(t)\right|^2\right\rangle=\frac{1}{\DD^2}\left\langle
 \sum_{jk}\rme^{-\rmi t(\sigma_j-\sigma_k)}
 \right\rangle.
 \end{split}
 \label{eq:sigma-sff-definition}
\end{equation}
This is the singular form factor of
Ref.~\cite{roccati_diagnosing_2024}, here applied to the centered
Liouvillian block. Defining the coherent singular state
\begin{equation}
 |\psi_\sigma\vket=\frac{1}{\sqrt{\DD}}
 \sum_{j=1}^{\DD}|v_j\vket,
 \qquad
 \sSFF(t)=\left\langle\left|
 \vbra\psi_\sigma|\rme^{-\rmi t\sqrt{\overline{\LL}^{\dagger}\overline{\LL}}}|\psi_\sigma\vket
 \right|^2\right\rangle,
 \label{eq:sigma-sff-return}
\end{equation}
shows that the spectral-sum and return-probability representations become equivalent again because $\sqrt{\overline{\LL}^{\dagger}\overline{\LL}}$ is Hermitian~\cite{roccati_diagnosing_2024}. 
This evolution is auxiliary: $\rme^{-\rmi t\sqrt{\overline{\LL}^{\dagger}\overline{\LL}}}$ is neither the physical propagator $\rme^{\LL t}$ nor, in general, a CPTP map, and $|\psi_\sigma\vket$ need not represent a physical density matrix.
For comparison with RMT, the main text figures show the connected form factor
after unfolding each realization. If $\widetilde{\sigma}_j$,
$j=1,\ldots,M$, are the retained unfolded singular values, we use
\begin{equation}
 K_{\sigma,c}(t)=\frac{1}{M}\left[
 \left\langle\left|\widetilde Z_\sigma(t)\right|^2\right\rangle
 -\left|\left\langle\widetilde Z_\sigma(t)\right\rangle\right|^2
 \right],
 \qquad
 \widetilde Z_\sigma(t)
 =\sum_{j=1}^{M}\rme^{-\rmi t\widetilde{\sigma}_j}.
 \label{eq:sigma-sff-connected}
\end{equation}
This normalization gives a unit plateau for a nondegenerate unfolded spectrum, while exact singular-value multiplicities modify it, as discussed for the integrable case in the End Matter.

\subsection*{B. Unfolding procedure}
\label{sec:unfolding}

\subsubsection*{1. Motivation}

Universal RMT predictions for spectral form factors describe fluctuations about the smooth spectral density, so quantitative comparison requires removing system-specific density variations by unfolding the spectrum to a locally uniform mean spacing~\cite{Guhr1998,haake_quantum_2018}. Otherwise, the local Heisenberg scale varies across the spectrum and nonuniversal density variations can obscure the universal ramp-to-plateau structure. For long-range observables such as the SFF, the unfolding prescription must therefore be controlled carefully~\cite{JiaVerbaarschot2020}. Since singular values form an ordered one-dimensional spectrum whereas Liouvillian eigenvalues form a two-dimensional point process, we use distinct unfolding procedures for the $\sSFF$ and $\DSFF$, with the latter following the conformal-unfolding and filtering approach of Ref.~\cite{LiYanProsenChan2024}.

\subsubsection*{2. Singular-value unfolding and normalization}
\label{app:sigma-conventions}

For the ordered singular-value spectrum, we use the standard
one-dimensional cumulative unfolding~\cite{Guhr1998}. For each realization
\(r\), we define the cumulative level-counting function
\begin{equation}
 {\cal N}_r(\sigma)
 =
 \sum_j\Theta(\sigma-\sigma_{rj})
 =
 \overline{\cal N}_r(\sigma)+{\cal N}_{r,\mathrm{fl}}(\sigma),
 \label{eq:sigma-staircase}
\end{equation}
where \(\overline{\cal N}_r(\sigma)\) is its smooth part. The unfolded
singular values are then
\begin{equation}
 \xi_{rj}=\overline{\cal N}_r(\sigma_{rj}),
\end{equation}
and therefore have unit mean density. 
Since the universal RMT form factor describes bulk correlations, we exclude the spectral edges before unfolding. In particular, the singular-value spectrum has a hard edge at \(\sigma=0\), where the two branches of the folded spectrum meet and the bulk folded GOE description no longer applies, whereas the upper edge is affected by the rapidly varying density and stronger finite-size corrections~\cite{Mehta2004,haake_quantum_2018, kawabata_singular-value_2023}. The precise extent of the excluded region is a numerical choice; throughout, we retain the central \(5\%\)--\(95\%\) rank window. For each realization \(r\), we first discard singular values \(\sigma\leq\epsilon_0=10^{-8}\) and denote by \(K_r\) the number of remaining values. After sorting them, we retain the indices \begin{equation} j_{\rm lo}=\left\lceil0.05K_r\right\rceil, \qquad j_{\rm hi}=\left\lfloor0.95K_r\right\rfloor, \qquad M=j_{\rm hi}-j_{\rm lo}+1\,. \label{eq:sigma-bulk-window} \end{equation} 
The same number \(M\) of bulk singular values is retained in every realization, so that all realizations enter the connected estimator with the same normalization. For the spectra used in Figs.~\ref{fig:single-mode-spectra} and \ref{fig:multimode-sigma-sff} with $N=10$ and $Q=4$, \(K_r=44100\) for every realization and \(M=39691\).
We sample the counting function over the retained window at \(41\) equally spaced knots between its smallest and largest retained singular values and fit the sampled counts by a least-squares polynomial \(p_r\).
To accommodate degenerate or insufficiently sampled spectra, its degree is chosen adaptively as
\begin{equation}
 d_r
 =
 \max\!\left[
 0,\min\!\left(5,n_{r,\mathrm{grid}}-1,n_{r,\mathrm{uniq}}-1\right)
 \right],
 \qquad
 \xi_{rj}=p_r(\sigma_{rj}),
 \label{eq:sigma-unfold-map}
\end{equation}
where \(n_{r,\mathrm{grid}}\) and \(n_{r,\mathrm{uniq}}\) denote, respectively, the numbers of valid sampled points and distinct abscissae available to the fit.
Writing \(\widetilde t\) for time in unfolded units, the trace and the normalized connected estimator are
\begin{equation}
\begin{split}
 Z_r(\widetilde t)
 =
 \sum_{j=1}^{M}
 e^{-2\pi\rmi\widetilde t\,\xi_{rj}},
 \qquad
 {\cal K}_{\sigma,c}(\widetilde t)
 =
 \frac{R}{R-1}\frac{1}{M}
 \left[
 \frac{1}{R}\sum_{r=1}^{R}|Z_r(\widetilde t)|^2
 -
 \left|
 \frac{1}{R}\sum_{r=1}^{R}Z_r(\widetilde t)
 \right|^2
 \right].
\end{split}
\label{eq:sigma-connected}
\end{equation}
The factor \(R/(R-1)\) removes the finite-ensemble covariance bias, while
division by \(M\) normalizes the nondegenerate diagonal contribution to
unity. In the main text, we denote
\({\cal K}_{\sigma,c}=K_c/K_\infty\) and use the normalization
\(K_\infty\equiv M\). Exact singular-value multiplicities can raise the true
late-time value above unity, as discussed for the integrable corner in the
End Matter.
The curves of Figs.~\ref{fig:atomic_spontaneous_emission} and~\ref{fig:cavity_dissipation} use \(R=16\) disorder realizations and \(400\)
logarithmically spaced points over
\(10^{-3}\leq\widetilde t\leq10\). For display only, they are also smoothed with an \(11\)-point,
degree-three Savitzky--Golay filter applied to the base-ten logarithm of the
ordinate.

\subsubsection*{3. Complex-eigenvalue unfolding, filtering, and calibration}
\label{app:complex-dsff-processing}

We adopt the conformal-unfolding and Gaussian-filtering procedure of Ref.~\cite{LiYanProsenChan2024}, followed by a separate calibration of the horizontal scale. 
For each realization, we remove only the steady eigenvalue, identified by \(|\lambda|\leq10^{-8}\), while retaining the real-axis modes and both members of every conjugate pair. 
From the two-dimensional eigenvalue density, averaged over all disorder realizations before unfolding, we locate its maximum \(\lambda_0\) on the real axis and apply the same conformal map to every realization, 
\begin{equation}
    \widetilde \lambda=-\rmi(\lambda-\lambda_0)^{1/2},
    \label{eq:complex-unfolding-map}
\end{equation}
with branch cut \([\lambda_0,\infty)\). This transformation corresponds to \(A=-\rmi\) and \(\beta=1/2\) in the notation of Ref.~\cite{LiYanProsenChan2024} and reduces the dominant density variation near the spectral edge.
From the ensemble-averaged density of the mapped spectrum, we determine its maximum \(\boldsymbol\mu=(\mu_x,\mu_y)\) and its Cartesian half-widths at half maximum \(\Delta_x\) and \(\Delta_y\). 
We then apply the Gaussian filter
\begin{equation}
     f(\widetilde \lambda)
     =
     \exp\!\left[
     -\alpha_x(\Real\widetilde \lambda-\mu_x)^2
     -\alpha_y(\Imag\widetilde \lambda-\mu_y)^2
     \right],
     \qquad
     \alpha_m=\Delta_m^{-2},
     \quad m=x,y\,,
    \label{eq:complex-dsff-filter}
\end{equation}
corresponding to unit dimensionless filter strength. For realization \(r\) and \(\phi=\pi/4\), we define the filtered spectral sum
\begin{equation}
     Z_r(|\tau|)
     =
     \sum_n f_{rn}
     \exp\!\left[
     \rmi|\tau|
     \left(
     \Real\widetilde \lambda_{rn}\cos\phi
     +
     \Imag\widetilde \lambda_{rn}\sin\phi
     \right)
     \right],
     \qquad
     f_{rn}=f(\widetilde \lambda_{rn}).
\end{equation}
The connected DSFF is normalized as
\begin{equation}
     \frac{K_c(|\tau|)}{P}
     =
     \frac{R}{R-1}\frac{1}{P}
     \left[
     \frac{1}{R}\sum_{r=1}^{R}|Z_r(|\tau|)|^2
     -
     \left|
     \frac{1}{R}\sum_{r=1}^{R}Z_r(|\tau|)
     \right|^2
     \right],
     \qquad
     P=\frac{1}{R}\sum_{r,n}f_{rn}^2.
     \label{eq:complex-dsff-production}
\end{equation}
Expanding \(|Z_r|^2\), the terms with identical eigenvalue indices (\(n=m\)) contribute \(\sum_n f_{rn}^2\), while the off-diagonal terms dephase at long times. 
Thus
\begin{equation}
    P=\frac{1}{R}\sum_{r,n}f_{rn}^2
\end{equation}
is the filter-weighted late-time plateau, and we normalize the connected DSFF by \(K_\infty=P\)~\cite{LiYanProsenChan2024}.
To compare the DSFF time scale across the different models and the
random-matrix reference, we define the time
scale directly from the local eigenvalue density within the filtered region. Assuming that the mapped density is approximately constant over the Gaussian
window, its value can be inferred from the filter-weighted plateau,
\begin{equation}
 P
 \simeq
 \rho_{\rm eff}\int_{\mathbb R^2}f^2\,\rmd^2\lambda
 =
 \frac{\pi\rho_{\rm eff}}
 {2\sqrt{\alpha_x\alpha_y}},
 \qquad
 \rho_{\rm eff}
 =
 \frac{2P\sqrt{\alpha_x\alpha_y}}{\pi},
 \qquad
 \widehat{\tau}
 =
 \frac{|\tau|}{\sqrt{\rho_{\rm eff}}}.
 \label{eq:complex-dsff-effective-density}
\end{equation}
We use \(\widehat{\tau}\), rather than \(\kappa\), for this dimensionless
coordinate to avoid confusion with the cavity-loss rate.
We calibrate the conversion between \(\widehat{\tau}\) and the bulk
Heisenberg scale using Gaussian complex-symmetric matrices in class
AI\(^{\dagger}\)~\cite{garcia-garcia_universality_2023},
\begin{equation}
 X_N=\frac{A+A^{\mathsf T}}{\sqrt2},
 \qquad
 A_{jk}=\frac{u_{jk}+\rmi v_{jk}}{\sqrt{2N}},
 \qquad
 u_{jk},v_{jk}\sim{\cal N}(0,1).
 \label{eq:ai-dagger-reference}
\end{equation}
We use \(N=128,256,512\), with \(512,256,128\) realizations, respectively.
The corresponding connected DSFFs are evaluated with
\(f(z)=\rme^{-|\lambda|^2}\) and the normalization in
Eq.~\eqref{eq:complex-dsff-production}. To estimate the large-\(N\) AI\(^{\dagger}\) reference we first interpolate
the curves for \(N=128,256,512\) onto the same \(\widehat{\tau}\) values.
At each \(\widehat{\tau}\), we fit the normalized DSFF linearly in \(1/N\)
and take the \(1/N\to0\) intercept, obtaining an estimate
\(F_{\mathrm{AI}^{\dagger}}(\widehat{\tau})\) of the infinite-size curve.
To determine its characteristic ramp-to-plateau time robustly in the
presence of statistical fluctuations, we fit the portion of this curve
after the correlation-hole minimum with a monotonically increasing
isotonic regression. We then define
\begin{equation}
 F_{\mathrm{AI}^{\dagger}}
 \!\left(\chi_{\mathrm{AI}^{\dagger}}\right)
 =0.95,
 \qquad
 \chi_{\mathrm{AI}^{\dagger}}=7.16.
 \label{eq:ai-dagger-calibration}
\end{equation}
and: 
\begin{equation}
 \tau_{\rm Hei}
 =
 \chi_{\mathrm{AI}^{\dagger}}\sqrt{\rho_{\rm eff}},
 \qquad
 \frac{|\tau|}{\tau_{\rm Hei}}
 =
 \frac{\widehat{\tau}}
 {\chi_{\mathrm{AI}^{\dagger}}}.
 \label{eq:complex-dsff-horizontal-scale}
\end{equation}
This calibration defines the common horizontal scale
\(|\tau|/\tau_{\rm Hei}\) used for the DSFF curve in the main text.

\subsection*{C. Bare spectral form factors}

\begin{figure}[t!]
 \centering
 \includegraphics[width=0.9\textwidth]{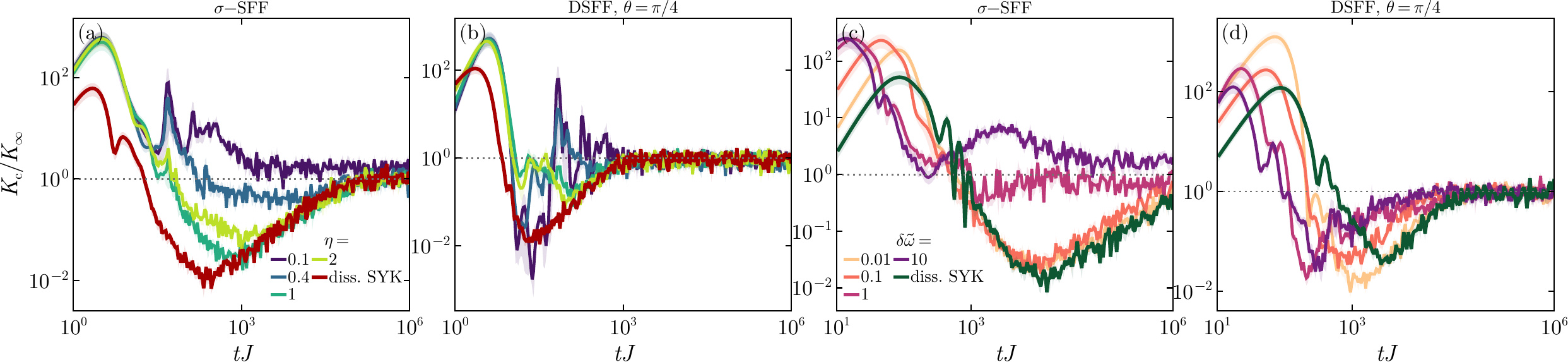}
 \caption{\label{fig:raw-sff}
 Bare connected $\sSFF$ and $\DSFF$ before unfolding for the single-mode model of Fig.~\ref{fig:atomic_spontaneous_emission} (panels (a) and (b)) and multimode model (panels c and d) of Fig.~\ref{fig:cavity_dissipation}, together with the dissipative $\SYK$ reference model.}
\end{figure}

Figure~\ref{fig:raw-sff} shows the connected $\sSFF$ and $\DSFF$ evaluated directly from the bare singular-value and Liouvillian spectra, before applying the unfolding and filtering procedures described in Sec.~\ref{sec:unfolding}. The integrability-to-chaos crossover remains qualitatively visible, while the curves also retain pronounced nonuniversal structure associated with the global spectral density and the corresponding model-dependent time scales. The unfolding procedures used in the main text remove these smooth spectral variations and isolate the universal bulk correlations required for a quantitative comparison with the random-matrix predictions.

\subsection*{D. Return probability of a coherent Gibbs state}

\begin{figure}[ht]
 \centering
 \includegraphics[width=0.7\textwidth]{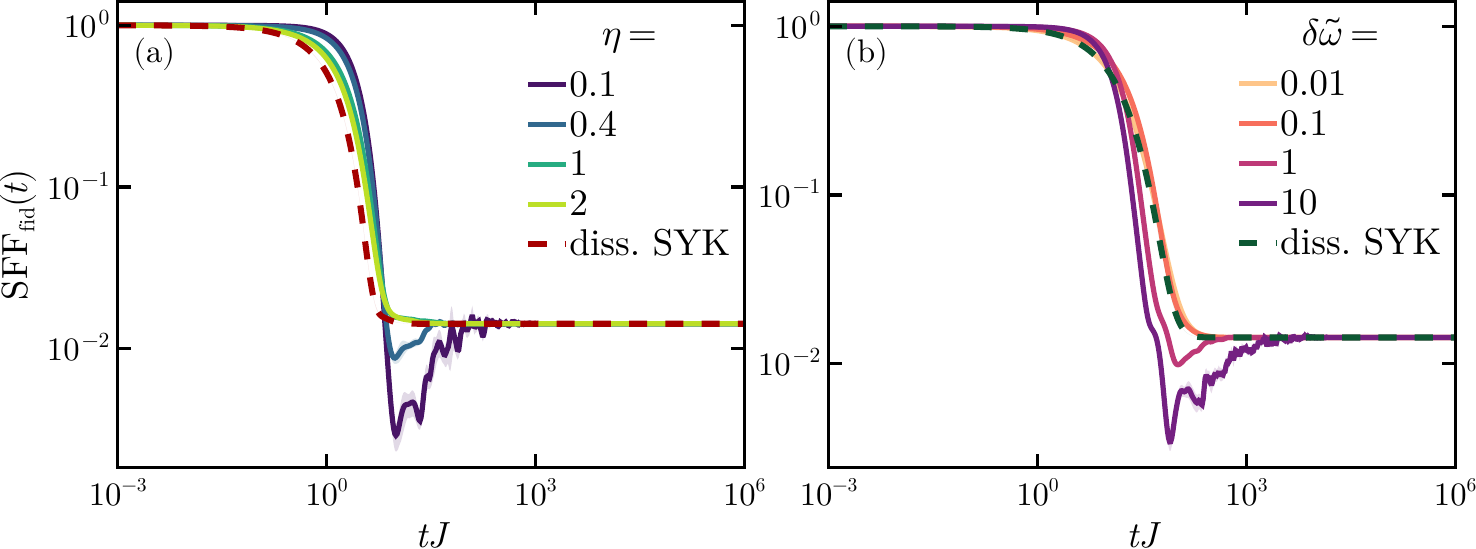}
 \caption{\label{fig:cgs-sff}
 Coherent-Gibbs-state return probability $\SFF_{\rm fid}(t)$ of Eq.~\ref{eq:fidelity-sff} for the single-mode model of Fig.~\ref{fig:atomic_spontaneous_emission} (panel a) and multimode model of Fig.~\ref{fig:cavity_dissipation} (panel b), together with the dissipative $\SYK$ reference model.}
\end{figure}

We illustrate the distinction stemming from the different continuations of the $\SFF$ to open systems discussed in Sec.~\ref{sec:sff-continuations} by evaluating the coherent-Gibbs-state return probability $\SFF_{\rm fid}(t)$ of Eq.~\eqref{eq:fidelity-sff}. Figure~\ref{fig:cgs-sff} shows this quantity for the same single-mode and multimode models considered in the main text, with $N=8$ and $Q=4$. In contrast to the $\DSFF$ and $\sSFF$, the return probability is dominated by the dissipative relaxation generated by $\rme^{\LL t}$
 and does not display the correlation hole associated with the onset of DQC. The spectral rigidity of the Liouvillian can therefore become manifest in the $\sSFF$ or $\DSFF$ while remaining hidden when using this definition.

\end{document}